\documentclass[11pt]{article}
\usepackage{amssymb,amsxtra}
\usepackage{graphicx}
\usepackage{amscd}
\usepackage{amsmath}
\usepackage{amsfonts}
\usepackage{amssymb}
\usepackage{amsthm}

\newtheorem{theorem}{Theorem}[section]

\newtheorem{corollary}[theorem]{Corollary}

\newtheorem{definition}[theorem]{Definition}

\newtheorem{remark}[theorem]{Remark}

\numberwithin{equation}{section}

\begin{document}
\title{\textbf{Riemann Zeroes and Phase Transitions via the Spectral Operator on Fractal Strings}\\}
\author{Hafedh Herichi and Michel\,L.\,Lapidus\,\footnote{The work of M.\,L.\,Lapidus was partially supported by the US National Science foundation under the research grants DMS-0707524 and DMS-1107750.}}
\maketitle

\begin{abstract}
The spectral operator was introduced by M.\,L.\,Lapidus and M.\,van Frankenhuijsen \cite{La-vF3} in their reinterpretation of the earlier work of M.\,L.\,Lapidus and H.\,Maier \cite{LaMa2} on inverse spectral problems and the Riemann hypothesis.\,In essence,\,it is a map that sends the geometry of a fractal string onto its spectrum.\,In this survey paper,\,we present the rigorous functional analytic framework given by the authors in \cite{HerLa1} and within which to study the spectral operator.\,Furthermore, we give a necessary and sufficient condition for the invertibility of the spectral operator (in the critical strip) and therefore obtain a new spectral and operator-theoretic reformulation of the Riemann hypothesis.\,More specifically,\,we show that the spectral operator is invertible (or equivalently,\,that zero does not belong to its spectrum) if and only if the Riemann zeta function $\zeta(s)$ does not have any zeroes on the vertical line $Re(s)=c$.\,Hence,\,it is not invertible in the mid-fractal case when $c=\frac{1}{2}$,\,and it is invertible everywhere else (i.e.,\,for all $c\in(0,1)$\,with $c\neq\frac{1}{2}$) if and only if the Riemann hypothesis is true.\,We also show the existence of four types of (mathematical) phase transitions occurring for the spectral operator at the critical fractal dimension $c=\frac{1}{2}$ and $c=1$ concerning the shape of the spectrum, its boundedness, its invertibility as well as its quasi-invertibility.
\end{abstract}

\newpage
\begin{quotation}

\textbf{AMS Classification (2010)}:\,\emph{Primary} 11M06, 11M26, 11M41, 28A80, 32B40, 47A10, 47B25, 65N21, 81Q12, 82B27.\,\emph{Secondary} 11M55, 28A75, 34L05, 34L20, 35P20, 47B44, 47D03, 81R40.\\

\end{quotation}

\tableofcontents

\vspace*{4mm}
\begin{center}
\textbf{\,Introduction}
\end{center}

\hspace*{3mm}The \emph{Theory of Fractal Strings and their Complex Dimensions}, developed by M.\,L.\,Lapidus and M.\,van\,Frankenhuijsen, investigates \emph{the geometric and physical properties of fractals} \cite{La-vF2,La-vF3} and describes \emph{the oscillations in the geometry and the spectrum of fractal strings} \cite{La-vF1,La-vF2}.\,Such oscillations are encoded in the complex dimensions of the fractal string, which are defined as the poles of the corresponding geometric zeta function.\\

\hspace*{3mm}The theory of fractal strings and their complex dimensions has a variety of applications to number theory, arithmetic geometry, spectral geometry, dynamical systems, mathematical physics and noncommutative geometry; see \cite{La3, La-vF2, La-vF3, La-vF4, La5}.\,Moreover, along with its companion (the theory of `fractal membranes' or `quantized fractal strings', developed in \cite{La5}), the theory is connected in several different manners with (and has potential physical applications to) aspects of quantum mechanics, quantum statistical physics, renormalization theory and the theory of critical phenomena, quantum field theory and string theory; see especially \cite{La5}, along with \cite{La-vF3} and \cite{La3}.\,The physical applications of the mathematical theory have yet to be fully investigated and developed, however.\\

\hspace*{3mm}In \cite{LaMa2}, a spectral reformulation of the Riemann hypothesis was obtained by M.\,L.\,Lapidus and Helmut Maier, in terms of inverse spectral problems for fractal strings.\,In short, one can always hear whether a given fractal string of dimension $c$ (different from $\frac{1}{2}$) is Minkowski measurable if and only if the Riemann hypothesis is true.\,Later on, this work was revisited in light of the theory of complex dimensions of fractal strings developed in \cite{La-vF2, La-vF3}.\,Moreover, in \cite{La-vF3}, the \emph{spectral operator} $\mathfrak{a}$ was introduced semi-heuristically as the operator that sends the geometry of a fractal string onto its spectrum.\\

\hspace*{3mm}In our recent joint work \cite{HerLa1}, we provided a rigorous functional analytic framework for the study of the spectral operator $\mathfrak{a}$.\,We showed that $\mathfrak{a}$ is an unbounded normal operator acting on a suitable scale of Hilbert spaces (indexed by the Minkowski dimension $c$ in (0,1) of the underlying fractal string) and precisely determined its spectrum (which turns out to be equal to the closure of the range of values of the Riemann zeta function along the vertical line $Re(s)=c$).\,Furthermore, we introduced a suitable family of truncated spectral operators $\mathfrak{a}^{(T)}$ and deduced that for a given $c>0$, the spectral operator is quasi-invertible (i.e., each of the truncated spectral operators is invertible) if and only if there are no Riemann zeroes on the vertical line of equation  $Re(s)=c$.\,It follows that the associated inverse spectral problem has a positive answer for all possible dimensions $c\in (0,1)$, other than the mid-fractal case when $c=\frac{1}{2}$, if and only if the Riemann hypothesis is true.\,Using results concerning the universality of the Riemann zeta function among the class of non-vanishing holomorphic functions, we also showed that the spectral operator is invertible for $c>1$, not invertible for $\frac{1}{2}<c<1$, and conditionally (i.e., under the Riemann hypothesis), invertible for $0<c<\frac{1}{2}$.\\

\hspace*{3mm}Moreover, we proved that the spectrum of the spectral operator is bounded for $c>1$, unbounded for $c=1$, equals the entire complex plane for $\frac{1}{2}<c<1$, and unbounded but, conditionally, not the whole complex plane, for $0<c<\frac{1}{2}$.\,We therefore deduced that four types of (mathematical) phase transitions occur for the spectral operator at the critical values (or \emph{critical fractal dimensions}) $c=\frac{1}{2}$ and $c=1$, concerning the shape of its spectrum, its boundedness (the spectral operator is bounded for $c>1$, unbounded otherwise), its invertibility (with phase transitions at $c=1$ and, conditionally, at $c=\frac{1}{2}$), as well as its quasi-invertibility (with a single phase transition at $c=\frac{1}{2}$ if and only if the Riemann hypothesis holds true).\\

\hspace*{3mm}From a philosophical and physical point of view, these new developments allow us to provide a natural quantization of many aspects of analytic number theory; that is, to associate natural operators to L-functions (such as the Riemann zeta function) and to correspondly quantize (and extend) the identities they satisfy.\\

\hspace*{3mm}The goal of the present survey article is to present an overview of some aspects of the work in \cite{HerLa1} (and, to a lesser extent, \cite{HerLa2}), in a manner that is accessible to physicists and other scientists or mathematicians interested in the multiple connections between physics, fractal geometry and number theory.\\

\begin{center}
\section{Generalized Fractal Strings and Their Complex Dimensions} 
\end{center}
 
\hspace*{3mm}In fractal geometry, an \emph{ordinary fractal string} is a bounded open subset $\Omega$ of the real line.\,Such a set is a disjoint union of open intervals, the lengths of which form a sequence
 
\begin{equation}
\mathcal{L}=l_{1},l_{2},l_{3},...
\end{equation}
\text
which we will assume to be infinite and such that $\{l_{j}\}_{j\geq1}$ is a decreasing sequence of lengths.\,Important information about the geometry of $\mathcal{L}$ is contained in its \emph{geometric zeta function},

\begin{equation}
\zeta_{\mathcal{L}}(s)=\sum_{j=1}^{\infty}{l_{j}}^{s},
\end{equation}
\text
where $Re(s)>D_{\mathcal{L}}$.\,Here, $D_{\mathcal{L}}:=\inf\{\alpha\in \mathbb{R}:\,\sum_{j=1}^{\infty}l_{j}^{\alpha}<\infty\}$ is the dimension of $\mathcal{L}$; it is called the abscissa of convergence of the Dirichlet series $\sum_{j=1}^{\infty}{l_{j}}^{s}$ and coincides with the fractal (i.e., Minkowski or box) dimension of the boundary of $\Omega$.\,Furthermore, $\zeta_{\mathcal{L}}$ is assumed to have a suitable meromorphic extension to an appropriate domain of the complex plane.\\

\hspace*{3mm}\emph{The complex dimensions} of an ordinary fractal string $\mathcal{L}$ are defined as the poles of the meromorphic extension of $\zeta_{\mathcal{L}}$.\,Interesting information about the geometric, spectral (i.e., vibrational) and dynamical \emph{oscillations} of a fractal string is encoded in both the real parts and imaginary parts of its complex dimensions (see \cite{La-vF1,La-vF2,La-vF3,La-vF4} for more information about the theory of ordinary fractal strings and their complex dimensions; see also Remark \ref{Rk:com} below).\\

\begin{remark}
From a physical point of view, the \emph{real parts} of the complex dimensions correspond to the \emph{amplitudes} of the vibrations or oscillations $($the larger the real parts, the larger the amplitudes$)$, while the \emph{imaginary parts} correspond to the \emph{frequencies} of oscillations $($the larger the imaginary parts, the faster the vibrations$)$.\,Mathematically, this statement is fully justified by the explicit formulas of \emph{\cite{La-vF2,La-vF3}}, to be briefly discussed in Theorem \ref{Thm:2} and the text surrounding it.\,Returning to a physical and heuristic interpretation, one could think more generally that the oscillations corresponding to the complex dimensions take place in a \emph{space of scales} naturally associated with the object (here, the space of reciprocal lengths $l^{-1}_{j}$ of an ordinary fractal string).
\end{remark}

\begin{remark}\label{Rk:com}
The theory of fractal strings originated in \emph{[La1--4]}, \emph{[La-Po1--3]}, \emph{[LaMa1--2]} and in the memoir \emph{[HeLa]}.\,It was pursued in many directions since then, by the second author and his collaborators, while the mathematical theory of complex fractal dimensions developed and matured; see the books \emph{[La-vF2]}, \emph{[La-vF3]} and \emph{[La5]}.\,See, especially, the second revised and enlarged edition of \emph{[La-vF3]} for an overview of the theory and for a number of relevant references, including \emph{\cite{HamLa}} for the case of random fractal strings, as well as \emph{[LaPe]} and \emph{[LaPeWi]} where the beginning of a higher-dimensional theory of complex dimensions of fractals is developed, particularly under suitable assumptions of self-similarity.
\end{remark}

\hspace*{3mm}The \emph{Cantor string}, denoted by $CS$, and defined as the complement of the Cantor set in the closed unit interval $[0,1]$ is a standard example of an ordinary fractal string:\\

$CS=(\frac{1}{3},\frac{2}{3})\bigcup(\frac{1}{9},\frac{2}{9})\bigcup(\frac{7}{9},\frac{8}{9})\bigcup(\frac{1}{27},\frac{2}{27})\bigcup(\frac{7}{27},\frac{8}{27})\bigcup(\frac{19}{27},\frac{20}{27})\bigcup(\frac{25}{27},\frac{26}{27})\bigcup ...$\\

Here, each length $l_{j}=3^{-j-1}$, $j\geq1$ is counted with a multiplicity $w_{j}=2^{j}$.\,Thus, the geometric zeta function associated to such a string is 

\begin{equation}
\zeta_{CS}(s)=\sum_{j=1}^{\infty}2^{j}.3^{-(j+1)s}=\frac{3^{-s}}{1-2.3^{-s}}\notag\\
\end{equation}
\text
whose set of poles is the set of complex numbers\\
\begin{equation}
\mathcal{D}_{CS}=\{D+in\textbf{p}:\,n\in\mathbb{Z}\},\notag\\
\end{equation}
\text 
where $D=\log_{3}2$ is the dimension of the $CS$ and \textbf{p}=$\frac{2\pi}{\log 3}$.\,This set is called the set of complex dimensions of $CS$.\,Note that the real part of these complex numbers is the Minkowski dimension of $CS$ and that the imaginary parts corresponds to the oscillatory period \textbf{p} in \emph{the volume of the inner tubular neighborhoods}\footnote{For a given $\epsilon>0$, the volume of the inner tubular neighborhood of the boundary, $\partial\Omega$, of a fractal string $\mathcal{L}$ with radius $\epsilon>0$, is $V(\epsilon)=vol_{1}\{x\in\Omega:\,d(x,\partial\Omega)<\epsilon\}$, where $vol_{1}$ is the one-dimensional Lebesgue measure on $\mathbb{R}$.\,In the case of the Cantor string and as shown in \cite{La-vF3}, we have $V_{CS}(\epsilon)=\frac{2^{-D}\epsilon^{1-D}}{D(1-D)\log 3}+\frac{1}{\log 3}\sum_{n=1}^{\infty}Re(\frac{(2\epsilon)^{1-D-in\textbf{p}}}{D+in\textbf{p}})-2\epsilon$.} of $CS$.\\

\hspace*{3mm}In fractal geometry,\,the classical definition of fractality,\,as stated by Mandelbrot,\,is the following:\\

\begin{quotation}
\emph{\textquotedblleft A fractal is by definition a set for which the Hausdorff--Besicovitch dimension strictly exceeds the topological dimension\textquotedblright} (see \cite{Man}).
\end{quotation}

\hspace*{3mm}In the light of the theory of complex dimensions, 
\begin{quotation}
\emph{\textquotedblleft An object will be called fractal if its geometric zeta function has at least one complex dimension with positive real part\textquotedblright} (see \cite{La-vF3}).
\end{quotation}

\vspace*{4mm}
\hspace*{3mm}\,For instance, the Devil's staircase (that is, the Cantor curve) is an example of a set which is non-fractal, in the classical Mandelbrot sense of fractality.\footnote{Indeed, its topological, Hausdorff and Minkowski (or box) dimensions all coincide and are equal to one  because the Cantor curve has finite length.}\,One wishes to describe the Devil's staircase as a fractal due to its (non-trivial) self-similar (or rather, self-affine) geometric structure.\\

\hspace*{3mm}However, the theory of complex dimensions classifies this set as a fractal.\,Such a result is obtained by finding the poles of the associated geometric zeta function, infinitely many of which turned out to be complex numbers with a positive real part; namely, in addition to 1 (the box dimension of the curve), these are the same complex dimensions as for the Cantor string.\,(See [La-vF3, \S12.2.2].)\,The theory of complex dimensions recaptures and extends the definition of fractality given by Mandelbrot to other sets which one wishes to describe as fractals in the classical sense.\,It enables one to obtain a better and more detailed understanding of the notion of fractality of certain rough sets in mathematical analysis, number theory and geometry and as a result, should have fruitful applications to the physical sciences.\\

\hspace*{3mm}One can \emph{listen to the sound} of a given ordinary fractal string $\mathcal{L}=\{l_{j}\}_{j=1}^{\infty}$.\,Here, the positive numbers $l_{j}$ denote the lengths of the connected components (i.e., open intervals) of a bounded open set $\Omega$ of the real line $\mathbb{R}$, with (possibly) fractal boundary $\partial\Omega$.\,In fact, spectral information about $\mathcal{L}$ is encoded by its \emph{spectral zeta function}, which is defined as 

\begin{equation}
\zeta_{\mathcal{\nu}}(s)=\sum_{f} f^{-s},
\end{equation}
\text
where $f=kl_{j}^{-1}$ ($k,j=1,2,...$) are the normalized frequencies of $\mathcal{L}$.\,Up to a trivial normalization factor, these are simply the square roots of the eigenvalues of the Laplacian (or free Hamiltonian) on $\Omega$, with Dirichlet boundary conditions on $\partial\Omega$.\,So that, in particular, the associated eigenfunctions are constrained to have nodes at each of the endpoints of the intervals of which the open set $\Omega$ is composed (see [La1--5, LaPo1--3, LaMa1--2, HeLa, La-vF2, La-vF3] for more details).\,\emph{The geometry and the spectrum} of  $\mathcal{L}$\space are related via the following formula (see [La2--3,La-vF3]):
\begin{equation}
\zeta_{\mathcal{\nu}}(s) =\zeta_{\mathcal{L}}(s) .\zeta(s),\label{Eq:gs}
\end{equation}
where $\zeta$ is the Riemann zeta function.\,Here, $\zeta_{\mathcal{L}}$ is the \emph{geometric zeta function} of $\mathcal{L}$, defined by $\zeta_{\mathcal{L}}(s)=\sum_{j=1}^{\infty}l_{j}^{s}$, for $Re(s)>D_{\mathcal{L}}$, the \emph{abscissa of convergence} of the Dirichlet series $\sum_{j=1}^{\infty}l_{j}^{s}$ or \emph{dimension} of $\mathcal{L}$ (which coincides with the Minkowski dimension of $\partial\Omega$, see [La2, La-vF3]).\,In the sequel, following a well-established tradition, we still denote by $\zeta_{\mathcal{L}}$ the meromorphic continuation (when it exists) of the geometric zeta function.\,An analogous convention will be used for all the other zeta functions encountered throughout this paper.\\

\begin{remark}
Equation \emph{(\ref{Eq:gs})} has since been given an interesting dynamical interpretation by Alexander Teplyaev in \emph{\cite{Tep1, Tep2}} in terms of the complex dymanics of the renormalization map \emph{\cite{Ram, RamTo, Sh, FukSh, Sab1, Sab2, Sab3}} associated with the decimation method for the Dirichlet Laplacian on a `fractal interval' $($i.e., an interval viewed as a nontrivial self-similar set or graph$)$; see also Derfer et al.\,\emph{\cite{DerGrVo}}.

\end{remark}

\hspace*{3mm}In [La-vF2, La-vF3], a \emph{generalized fractal string} $\eta$ is defined as a local positive or complex measure on $(0,+\infty)$\,satisfying $|\eta|(0,x_{0})=0$,\,for some $x_{0}>0$.\,Here, the positive measure $|\eta|$ is the variation of $\eta$.\footnote{For an introduction to measure theory, we refer, e.g., to \cite{Coh,Fo}.}\,For instance, let $\mathcal{L}=\{l_{j}\}_{j=1}^{\infty}$ be an ordinary fractal string with multiplicities $w_{j}$.\,Then the measure associated to $\mathcal{L}$,\,defined as $\eta_{\mathcal{L}}=\sum_{j=1}^{\infty}w_{j}\delta_{\{l_{j}^{-1}\}}$, is an example of a generalized fractal string ($\delta_{\{x\}}$ is the Dirac delta measure or the unit point mass concentrated at $x>0$).\,Note that in the case of an ordinary fractal string, $w_{j}$ is always integral for any $j\geq 1$.\,However, in general, this multiplicity is not necessarily integral.\,For instance, the \emph{generalized Cantor string} $\eta_{CS}=\sum_{j=1}^{\infty}b^{j}\delta_{\{a^{j}\}}$, where $1<b<a$, is an example of a generalized fractal string for which $w_{j}=b^{j}$ is non-integral. It is also the measure associated to the non-ordinary fractal string $\mathcal{L}=\{a^{-j}\}_{j\geq1}$, with (typically) non-integral multiplicities $b^{-j}$, $j\geq1$.\,Therefore, this example justifies the use of the word \textquotedblleft generalized\textquotedblright for this class of fractal strings.\\

\hspace*{3mm} Let $\eta$ be a generalized fractal string.\,Its \emph{dimension} is 
\begin{equation}
D_{\eta}:=\inf\Big\{\sigma\in\mathbb{R}: \int_{0}^{\infty}x^{-\sigma}|\eta|(dx)<\infty\Big\}.
\end{equation}
The \emph{counting function} of $\eta$ is\footnote{More precisely, in order to obtain accurate \emph{pointwise} formulas, one sets $N_{\eta}(x)=\frac{1}{2}(\eta(0,x]+\eta[0,x))$.}
\begin{equation}
N_{\eta}(x):=\int_{0}^{x}\eta(dx)=\eta(0,x).
\end{equation}

The \emph{geometric zeta function} associated to $\eta$ is the Mellin transform of $\eta$.\,It is defined as
\begin{equation} 
\zeta_{\eta}(s):=\int_{0}^{+\infty}x^{-s}\eta(dx) \mbox{\quad for $Re(s)>D_{\eta}$},
\end{equation}
where $D_{\eta}$ is the dimension of $\eta$.\,We assume that $\zeta_{\eta}$ has a meromorphic extension to some suitable neighborhood $\mathcal{W}$ of the half-plane $\{Re(s)>D_{\eta}\}$ (see [La-vF3, \S5.3] for more details on how $\mathcal{W}$ is defined) and we define the set $\mathcal{D}_{\mathcal{L}}(\mathcal{W})$ of visible \emph{complex dimensions} of $\eta$ by

\begin{equation}
\mathcal{D}_{\mathcal{L}}(\mathcal{W}):=\{\omega\in\mathcal{W}:\zeta_{\eta} \mbox{\quad has a pole at $\omega$}\}. 
\end{equation}

\begin{remark}
Since $\zeta_{\mathcal{L}}$ is assumed to be meromorphic, $\mathcal{D}_{\mathcal{L}}(W)$ is a \emph{discrete} $($and hence, at most countable$)$ subset of $\mathbb{C}$.\,Furthermore, since $\zeta_{\mathcal{L}}$ is holomorphic for $Re(s)>D_{\mathcal{L}}$ $($because by definition of $D_{\mathcal{L}}$, its Dirichlet series $\sum_{j=1}^{\infty}l_{j}^{s}$ is absolutely convergent there$)$, all the complex dimensions $\omega$ of $\mathcal{L}$ satisfy $Re(\omega)\leq D_{\mathcal{L}}$.
\end{remark}

\hspace*{3mm}The \emph{spectral measure} $\nu$ associated to $\eta$ is defined by
\begin{equation}
\nu(A)=\sum_{k=1}^{\infty}\eta\left(\frac{A}{k}\right),
\end{equation} 
\text
for any bounded Borel set (or equivalently, interval) $A\subset(0,+\infty)$.\\

\hspace*{3mm}The \emph{spectral zeta function} associated to $\eta$ is the geometric zeta function of $\nu$.\,It is shown in [La-vF3, \S4.2] that
\begin{equation} 
\nu=\eta \ast h =\eta\ast(\underset{p\in\mathcal{P}}{\ast h_{p}}),
\end{equation}
where $\ast$\space is the multiplicative convolution of measures on the multiplicative half-line $(0,+\infty)$ of `\emph{reciprocal scales}', $\mathcal{P}$ denotes the set of (rational) primes ($\mathcal{P}=\{2, 3, 5, 7, 11, ...\}$), $h=\sum_{k=1}^{\infty}\delta_{\{k\}}$ is the \emph{harmonic generalized fractal string} (as above, $\delta_{\{.\}}$ is the Dirac delta measure) and $h_{p}=\sum_{k=1}^{\infty}\delta_{\{p^{k}\}}$ is the \emph{prime harmonic generalized fractal string} (see [La-vF3, \S4.2.1]).\\

As a result,\,we have 
\begin{equation}
\zeta_{\mathfrak{h}}(s)=\zeta_{\underset{ p\in\mathcal{P}}{\ast\mathfrak{h_{p}}}}(s)=\zeta(s)=\underset{ p\in\mathcal{P}}{\prod}\frac{1}{1-p^{-s}}=\underset{ p\in\mathcal{P}}{\prod}\zeta_{\mathfrak{h}_{p}}(s),
\end{equation}
where $Re(s)>1$.\\

The \emph{spectral zeta function} associated to $\nu$,\space which is the geometric zeta function of $\nu$,\,is related to $\zeta_{\eta}$\space via the following formula (which is the exact analog of Equation (\ref{Eq:gs})):

\begin{equation} 
\zeta_{\nu}(s)=\zeta_{\eta}(s).\zeta(s),
\end{equation}
where $\zeta$\space is the Riemann zeta function.\,This latter function is known to have an Euler product expansion given by the formula 
\begin{equation}
\zeta(s)=\prod_{p\in \mathcal{P}}(1-p^{-s})^{-1},\mbox{\quad for\,\,} Re(s)>1,\label{eq:Ep}
\end{equation}
\text
where, as before, p runs over all the set $\mathcal{P}$ of all prime numbers.\\

\hspace*{3mm}Riemann showed in his celebrated 1858 paper on the distribution of prime numbers that $\zeta$ has a meromorphic continuation to all of $\mathbb{C}$ with a single (and simple) pole at $s=1$, which satisfies the \emph{functional equation}
\begin{equation}
\xi(s)=\xi(1-s),\mbox{\,}\,s\in \mathbb{C},\label{Eq:fE}
\end{equation}
where
\begin{equation}
\xi(s):=\pi^{-\frac{s}{2}}\Gamma(\frac{s}{2})\zeta(s)\label{Eq:CZ}
\end{equation}
is the \emph{completed} (or \emph{global}) Riemann zeta function.\,He also conjectured that the nontrivial (or \emph{critical}) zeroes of $\zeta(s)$ (i.e., the zeroes of $\zeta(s)$ which are in the \emph{critical strip} $0<Re(s)<1$) all lie on the \emph{critical line} $Re(s)=\frac{1}{2}$.\,This famous conjecture is known as the \emph{Riemann Hypothesis}.

\begin{remark}\label{Rk:EP}
It follows from the Euler product in Equation \emph{(\ref{eq:Ep})} that $\zeta(s)$ does not have any zeroes for $Re(s)>1$.\,According to the functional equation \emph{(\ref{Eq:fE})}, the same is true for $Re(s)<0$, except for the \emph{trivial zeroes} of $\zeta(s)$ at $s=-2n$, for $n=1,2,3,...$, which correspond to the poles of the gamma function $\Gamma(\frac{s}{2})$ in the definition \emph{(\ref{Eq:CZ})} of $\xi(s)$.\,Hadamard showed in 1892 that $\zeta(s)$ does not have any zeroes on the vertical line $Re(s)=1$, which was a key step in the proof of the Prime Number Theorem $($1896$)$.\,As a result, and in light of the functional equation \emph{(\ref{Eq:fE})}, $\zeta(s)$ does not have any zeroes either on the vertical line $Re(s)=0$.\,This is why the nontrivial zeroes of $\zeta(s)$ lie within the critical strip $0<Re(s)<1$ and are called the critical zeroes.\,$($Note that the zeroes of $\xi(s)$ coincide with the critical zeroes of $\zeta(s)$.$)$
\end{remark} 

\begin{remark}\label{Rk:rm}
In his same 1858 paper, Riemann obtained \emph{explicit formulas} connecting an expression involving the prime numbers $($for example, the prime number counting function$)$, on the one hand, and the $($trivial and critical$)$ zeroes of the Riemann zeta function $\zeta(s)$, on the other hand.\,This duality between the primes $p$ $($or their logarithms $\log p$$)$ and the zeroes $($and the pole$)$ of $\zeta(s)$ has been key to most approaches to the Riemann hypothesis.\,In a similar spirit, the generalization of Riemann's explicit formula obtained in \emph{(\cite{La-vF2,La-vF3})} and to be discussed next connects certain expressions involving a generalized fractal string $\eta$ $($for example, the geometric or the spectral counting function of $\eta$$)$, on the one hand, and the geometric or spectral complex dimensions of $\eta$, on the other hand; that is, the poles of the geometric or spectral zeta function of $\eta$.\footnote{Note that the zeroes and the pole of $\zeta$ (along with their multiplicities$)$ can be recovered from the poles $($and the sign of the residues$)$ of the logarithm derivative $-\frac{\zeta'(s)}{\zeta(s)}$.}
\end{remark}

Moreover, the second author and M.\,van Frankenhuijsen obained \emph{explicit distributional formulas} associated to $\eta$.\,In these formulas,\,the kth distributional primitive of $\eta$\space will be viewed as a \emph{distribution},\,acting on functions in Schwartz space \cite{Schw}.\,(See [La-vF3, Chapter 5] for a detailed discussion and a precise statement of these explicit formulas, both in their distributional and pointwise form.)

\begin{theorem}\emph{\cite{La-vF2,La-vF3}} \label{Thm:2}
Let $\eta$\,be a \emph{languid\footnote{A generalized fractal string $\eta$ is said to be \emph{languid} if its geometric zeta function $\zeta_{\eta}$ satisfies some suitable \emph{polynomial growth conditions}; see [La-vF3, \S5.3].} generalized fractal string}.\,Then, for any $k\in\mathbb{Z}$, its  kth distributional primitive is given by
\begin{align}
\mathcal{P}_{\eta}^{[k]}(x)=&\sum_{\omega\in \mathcal{D}_{\eta}(\mathcal{W})}res\left(\frac{x^{s+k-1}\zeta_{\eta}(s)}{(s)_{k}};\omega\right)+\frac{1}{(k-1)!}\sum_{j=0}^{k-1}C_{j}^{k-1}(-1)^{j}x^{k-1-j}\notag\\
                           &.\zeta_{\eta}(-j)+\mathcal{R}_{\eta}^{[k]}(x),
\end{align}                             
where $\omega$ runs through the set $\mathcal{D}_{\mathcal{L}}(W)$ of visible complex dimensions of $\eta$ and $\mathcal{R}_{\eta}^{[k]}(x)=\frac{1}{2\pi i}\int_{\mathcal{S}}x^{s+k-1}\zeta_{\eta}(s)\frac{ds}{(s)_{k}}$\,is the error term, which can be suitably estimated as $x \to +\infty$ and which, under appropriate hypotheses, vanishes identically $($thereby yielding \emph{exact} explicit formulas, see \emph{[La-vF3, \S5.3]}$)$.\footnote{Here, the coefficients $C^{k-1}_{j}$ are scalars defined in terms of the gamma function.\,Their exact expression is not important for the purpose of this paper.\,Moreover, the Pochhammer symbol is defined by $(s)_{k}=s(s+1)...(s+k-1)$, for $k\geq1$, and $(s)_{k}=\frac{\Gamma(s+k)}{\Gamma(s)}$ for any $k\in \mathbb{Z}$.}
\end{theorem}

\hspace*{3mm}The action of $\mathcal{P}_{\eta}^{[k]}$ on a test function $\varphi$ in Schwartz space is given by (with $\widetilde{\varphi}$ denoting the Mellin transform of $\varphi$)\\
\begin{align}
<\mathcal{P}_{\eta}^{[k]},\varphi>=&\sum_{\omega\in \mathcal{D}_{\eta}(\mathcal{W})}res\left(\frac{\zeta_{\eta}(s)\widetilde{\varphi}(s+k)}{(s)_{k}};\omega\right)+\frac{1}{(k-1)!}\sum_{j=0}^{k-1}C_{j}^{k-1}(-1)^{j}\notag\\
                                  &.\zeta_{\eta}(-j)\widetilde{\varphi}(k-j)+\mathcal{R}_{\eta}^{[k]}(x).
\end{align}

\hspace*{3mm}Recall from our earlier discussion (in Remark \ref{Rk:rm}) that the original explicit formula was first obtained by Riemann in 1858 as an analytical tool to understand the distribution of the primes.\,It was later extended by von Mangoldt and led in 1896 to the first rigorous proof of the Prime Number Theorem,\,independently by Hadamard and de la Vall\'ee Poussin.\,(See \cite{Edw, Ing, Ivi, Tit, Pat}.)\,In [La-vF3, \S5.5], the interested reader can find a discussion of how to recover the Prime Number Theorem, along with Riemann's original explicit formula and its various number theoretic extensions, from Theorem \ref{Thm:2} (and more general results given in [La-vF3, Chapter 5]).\\

\hspace*{3mm}While defined at level $k=0$, the distributional formula associated to $\eta$, which is provided in Theorem \ref{Thm:2}, gives an explicit expression for $\eta$ as a sum over the complex dimensions associated to $\eta$.\,This formula is similar to the formula of \emph{density of geometric states} (or \emph{density of scales} formula)\,(see [La-vF3,$\S$\,6.3.1]):\footnote{For clarity, we give these explicit formulas in the case of simple poles.}\\
\begin{equation}
\eta=\sum_{\,\omega\in\mathcal{D_{\eta}(W)\,}}res(\zeta_{\eta}(s);\omega)x^{\omega-1}.\label{Eq:18}
\end{equation}
When these formulas are applied to the spectral measure $\nu=\eta\ast h$,\,we obtain an explicit expression for $\nu$ as a sum over its complex dimensions which is similar to the formula of \emph{density of spectral states} in \emph{quantum physics} or (\emph{density of frequencies}\,formula):\\
\begin{align}
\nu
&=\zeta_{\eta}(1)+\sum_{\,\omega\in\mathcal{D_{\eta}(W)\,}}res(\zeta_{\eta}(s)\zeta(s)x^{s-1};\omega)\notag\\
&=\zeta_{\eta}(1)+\sum_{\omega\in \mathcal{D}_{\eta}(W)}res(\zeta_{\eta}(s);\omega)\zeta(\omega)x^{\omega-1},\label{Eq:19}
\end{align}
\text
where $1\notin \mathcal{D}_{\eta}(\mathcal{W})$ and, as above, $res(\zeta_{\eta}(s);\omega)$ denotes the residue of $\zeta_{\eta}(s)$ as $s=\omega$.\\

\hspace*{3mm}Next, we define \emph{the spectral operator} and present some of its fundamental properties, which were studied in \cite{HerLa1}. 
\begin{center}
\section{The Spectral Operator $\mathfrak{a}_{c}$ and the Infinitesimal Shift $\partial_{c}$} 
\end{center}

\hspace*{3mm}In [La-vF3, \S6.3.2], using the distributional formulas given in Theorem \ref{Thm:2} at level k=0 (see Equations (\ref{Eq:18}) and (\ref{Eq:19}) above), \emph{the spectral operator} was heuristically defined as \emph{the operator mapping the density of geometric states $\eta$\space to the density of spectral states $\nu$}:\\
\begin{equation}
\eta\longmapsto\nu
\end{equation}
\hspace*{3mm}At level k=1,\,it will be defined, on a suitable Hilbert space $\mathbb{H}_{c}$, where $c\geq 0$, as the operator mapping the counting function of $\eta$ to the counting function of $\nu=\eta\ast h$ (that is, mapping \emph{the geometric counting function} $N_{\eta}$ onto the \emph{spectral counting function} $N_{\nu}$):\footnote{In the sequel, unless explicitly stated oherwise, the symbol $k\in \mathbb{N}$ will be used as a dummy index rather than to refer to the level $k=0$ or $k=1$ at which Theorem \ref{Thm:2} is applied.}
\begin{equation}
N_{\eta}(x)\longmapsto\nu(N_{\eta})(x)=N_{\nu}(x)=\sum_{k=1}^{\infty}N_{\eta}\left(\frac{x}{k}\right).
\end{equation}

\hspace*{3mm}Note that under the change of variable $x=e^{t}$, where $t\in \mathbb{R}$ and $x>0$, one can obtain an \emph{additive} representation of the spectral operator $\mathfrak{a}$
\begin{equation}\label{Eq:spop}
f(t)\mapsto \mathfrak{a}(f)(t)=\sum_{k=1}^{\infty}f(t-\log k),
\end{equation}
and of its \emph{operator-valued Euler factors} $\mathfrak{a}_{p}$
\begin{equation}
f(t)\mapsto \mathfrak{a}_{p}(f)(t)=\sum_{k=0}^{\infty}f(t-k\log p).
\end{equation}
These operators are related by an \emph{Euler product} (here, the product is the composition of operators) as follows:

\begin{equation}
f(t)\mapsto \mathfrak{a}(f)(t)=\left(\prod_{p\in \mathcal{P}}\mathfrak{a}_{p}\right)(f)(t).\label{Eq:EPr}
\end{equation} 

\hspace*{3mm}Let $f$ be an infinitely differentiable function on $\mathbb{R}$.\footnote{In our later, more mathematical discussion, $f$ will not necessarily be the counting function of some generalized fractal string $\eta$, but will instead be allowed to be an element of the Hilbert space $\mathbb{H}_{c}$ (with possibly some additional conditions on $f$ or on the parameter $c$);  see Equation (\ref{Eq:HS}) below and the text surrounding it, along with Equations (\ref{Eq:acf}) and (\ref{Eq:Dsp}).} \,Then, the Taylor series of $f$ can be written as
\begin{align}
f(t+h)&=f(t)+\frac{hf'(t)}{1!}+\frac{h^{2}f^{''}(t)}{2!}+...\notag\\
      &=e^{h\frac{d}{dt}}(f)(t)=e^{h\partial}(f)(t),\notag
\end{align}       
where $\partial=\frac{d}{dt}$ is the first order differential operator with respect to $t$.\footnote{This differential operator is also the infinitesimal generator of the (one-parameter) group of shifts on the real line.\,For this reason, it is also called the \emph{infinitesimal shift}; see Remark \ref{Rk:2.15}.}\,Note that this yields a new \emph{heuristic} representation for the spectral operator and its prime factors:

\begin{align}
\mathfrak{a}(f)(t)&=\sum_{k=1}^{\infty}e^{-(\log k)\partial}(f)(t)=\sum_{k=1}^{\infty}\left(\frac{1}{k^{\partial}}\right)(f)(t)\notag\\
                  &=\zeta(\partial)(f)(t)=\zeta_{h}(\partial)(f)(t)=\prod_{p\in \mathcal{P}}(1-p^{-\partial})^{-1}(f)(t)\label{Eq:Spr}
\end{align}  
and for any prime $p$,
\begin{align}
\mathfrak{a}_{p}(f)(t)&=\sum_{k=0}^{\infty}f(t-k\log p)=\sum_{k=0}^{\infty}e^{-k(\log p)\partial}(f)(t)=\sum_{k=0}^{\infty}\left(\frac{1}{p^{k}}\right)\partial(f)(t)\notag\\
                      &=\left(\frac{1}{1-p^{-\partial}}\right)(f)(t)=(1-p^{-\partial})^{-1}(f)(t)=\zeta_{h_{p}}(\partial)(t).\label{Eq:Pf}
\end{align}

\hspace*{3mm}The representations of $\mathfrak{a}$ and $\mathfrak{a}_{p}$, given  respectively in Equations (\ref{Eq:Spr}) and (\ref{Eq:Pf}) above, were provided without proof in [La-vF3, $\S$6.3.2] but have since been rigorously justified in \cite{HerLa1} and in its sequel \cite{HerLa2}.\\

\begin{remark}Note that at this stage of the exposition, we have not yet defined a Hilbert space, a core or even a domain allowing us to precisely define and study each of these operators.
\end{remark}

\hspace*{3mm}The map in Equation (\ref{Eq:spop}) \emph{relates the spectrum of a fractal string with its geometry}.\,The problem of deducing geometric information from the spectrum of a fractal string, or equivalently, of addressing the question 
\begin{center}
\begin{quotation}
\emph{\textquotedblleft Can one hear the shape of a fractal string?\textquotedblright},
\end{quotation}
\end{center}
was first studied by the second author and H.\,Maier \cite{LaMa1,LaMa2}.\,\emph{The inverse spectral problem} they considered was the following: \\

\textquotedblleft \emph{Given any fixed} $D\in (0,1)$, \emph{and any fractal string} $\mathcal{L}$ \emph{of dimension D such that for some constant} $c_{D}>0$ \emph{and} $\delta>0$, \emph{we have}\footnote{See Remark \ref{Rk:W} for the definition of the Weyl term, $W(x)$.}
\begin{equation}\label{Eq:Wt}
N_{\nu}(x)=W(x)-c_{D}x^{D}+ O(x^{D-\delta}),\mbox{\quad as\,\,} x\to +\infty,
\end{equation}
\emph{is it true that} $\mathcal{L}$ \emph{is Minkowski measurable}?\textquotedblright.\\

\hspace*{3mm}In particular, they showed that this inverse spectral problem has a positive answer for all fractal strings of dimension $D\in(0,1)-\frac{1}{2}$\,if and only if the Riemann hypothesis holds,\,i.e.,\,if and only if the Riemann zeta function $\zeta(s)$\,does not vanish for $Re(s)\ne\frac{1}{2},\,Re(s)>0$; see \cite{LaMa1,LaMa2}.\\

\hspace*{3mm}More specifically, they showed in \cite{LaMa2} that for a given $D\in (0,1)$, the above inverse spectral problem has an affirmative answer if and only if $\zeta(s)$ does not have any zeroes on the vertical line $Re(s)=D$.\,It follows that this problem has a negative answer in the mid-fractal case when $D=\frac{1}{2}$ (since $\zeta(s)$ has zeroes on the critical line $Re(s)=\frac{1}{2}$) but that it has a positive answer for all $D\in (0,1)$, $D\ne \frac{1}{2}$ (or equivalently, in light of the functional equation (\ref{Eq:fE}), for all $D\in (0,\frac{1}{2})$) if and only if the Riemann hypothesis is true.\,In \S3 below, based on the work in \cite{HerLa1}, we will give a precise operator theoretic interpretation of this result, as well as a suitable extension thereof, in terms of an appropriate notion of invertibility of the spectral operator $\mathfrak{a}$.

\begin{remark}
The work in \emph{\cite{LaMa1,LaMa2}} was revisited and extended to a large class of arithmetic zeta functions in \emph{\cite{La-vF1, La-vF2, La-vF3}}, using the explicit formulas recalled in Theorem \ref{Thm:2}.\,$($See \emph{[La-vF3, Chapter 9]}.$)$ Our work in \emph{\cite{HerLa1, HerLa2}} can also be extended to this more general setting, but we will not discuss this extension here, for the simplicity of exposition.
\end{remark}

\begin{remark}\label{Rk:W}
In the statement of the above inverse spectral problem, the Weyl term $W(x)$ is the leading asymptotic term.\,Namely,
\begin{equation}
W(x)=(2\pi)^{-1}|\Omega|x,
\end{equation}
where $x$ is the $($normalized$)$ frequency variable and $|\Omega|$ denotes the \textquotedblleft volume\textquotedblright\\$($really, the length$)$ of $\Omega\subset \mathbb{R}$.\\

\hspace*{3mm}Furthermore, recall that $\mathcal{L}$ $($or equivalently, $\partial\Omega$, the boundary of $\Omega$$)$ is said to be \emph{Minkowski measurable} if the following limit exists in $(0,+\infty):$

\begin{equation}
\lim_{\epsilon\to 0^{+}}\frac{|\Omega_{\epsilon}|}{\epsilon^{1-D}}:=\mathcal{M}(\mathcal{L}),\label{Eq:Mm}
\end{equation}
where $\mathcal{M}(\mathcal{L})$ is called the \emph{Minkowski content} of $\mathcal{L}$ $($or of $\partial{\Omega}$$)$ and $|\Omega_{\epsilon}|$ denotes the volume of the inner $\epsilon$-neighborhood of $\partial \Omega$:

\begin{equation}
\Omega_{\epsilon}=\{x\in \Omega:\,dis(x,\partial \Omega)<\epsilon\}.
\end{equation}
It then follows that $D$ is the Minkowski $($or box$)$ dimension of $\mathcal{L}$ $($i.e., of $\partial \Omega$$)$.
\end{remark}

\begin{remark}\label{Rk:2.4}
Prior to the work in \emph{\cite{LaMa1, LaMa2}}, the second author and Carl Pomerance \emph{\cite{LaPo1, LaPo2}} had studied the corresponding \emph{direct spectral problem} for fractal strings.\,More specifically, they had shown, in particular, that if $\mathcal{L}$ is Minkowski measurable $($which, according to a key result in \emph{\cite{LaPo1, LaPo2}}, is true iff $l_{j}\sim L.j^{-\frac{1}{D}}$ as $j\to +\infty$ or equivalently, iff $N_{\mathcal{L}}(x)\sim C.x^{D}$ as $x\to +\infty$, for some $L>0$ and $C>0$$)$, then the eigenvalue $($or rather, frequency$)$ counting function $N_{\nu}(x)$ satisfies Equation \emph{(\ref{Eq:Wt})} $($with $o(x^{D})$ instead of $O(x^{D-\delta})$$)$, where
\begin{equation}
c_{D}:=2^{-(1-D)}(1-D)(-\zeta(D))\mathcal{M}(\mathcal{L})
\end{equation}
\text
and $\mathcal{M}(\mathcal{L})$ is the Minkowski content of $\mathcal{L}$, as defined in Equation \emph{(\ref{Eq:Mm})}.\\$($Note that $-\zeta(D)>0$ for $0<D<1$.$)$ This result resolved in the affirmative the $($one-dimensional$)$ \emph{modified Weyl--Berry conjecture} $($as formulated in \emph{\cite{La1}}$)$.
\end{remark}

\begin{remark}
For results concerning the higher-dimensional analog of the above inverse and direct spectral problems for \textquotedblleft\emph{fractal drums}\textquotedblright in $\mathbb{R}^{d}$ $($$d\geq 1$$)$, as well as for their physical motivations and their relationship with the Weyl--Berry conjecture \emph{\cite{Berr1, Berr2}}, the reader may wish to consult \emph{\cite{La1, La2, La3, La4, LaPo3}}, along with \emph{[La-vF3, \S12.5]} and the references therein, including \emph{\cite{Berr1, Berr2, BroCa, FlVa, Ger, GerScm1, GerScm2, HeLa, vB-Gi}}.\,$($See also \emph{\cite{FukSh, Ham1, Ham2, Ki, KiLa1, KiLa2, La3, Sab3, Str, Tep1, Tep2}} and the relevant references therein for the case of a drum with a fractal membrane instead of a fractal boundary.$)$\\
\end{remark}

\hspace*{3mm}In \cite{HerLa1},  we start by providing a functional analytic framework enabling us to rigorously study the spectral operator.\,This functional analytic framework is based in part on defining a precise weighted Hilbert space $\mathbb{H}_{c}$, dependent on a parameter $c\geq 0$, in which the spectral operator is acting and then defining and studying this operator.\,We set

\begin{equation}\label{Eq:HS}
\mathbb{H}_{c}=L^{2}(\mathbb{R},\mu_{c}(dt)),
\end{equation}
\text
where $\mu_{c}(dt)=e^{-2ct}dt$ (here, $dt$ is the Lebesgue measure on $\mathbb{R}$).\\

\begin{remark}
Note that $\mathbb{H}_{c}$ is the space of Lebesgue square-integrable functions $f$ with respect to the positive \emph{weight function} $w(t)=e^{-2ct}:$

\begin{equation}\label{Eq:fc}
||f||^{2}_{c}=\int_{0}^{+\infty}|f(t)|^{2}e^{-2ct}dt<\infty.
\end{equation}
\text
It is obtained by completing the space $\mathcal{H}_{c}$ of infinitely differentiable functions $f$ on $\mathbb{R}=(-\infty,+\infty)$ satisfying the finiteness condition \emph{(\ref{Eq:fc})}.\,$($It follows, of course, that $\mathcal{H}_{c}$ is dense in $\mathbb{H}_{c}$.$)$
\end{remark}

The Hilbert space $\mathbb{H}_{c}$ is equipped with the following inner product
\[<f,g>_{c}\,=\int_{0}^{\infty}f(t)\overline{g(t)}e^{-2ct}dt\]
and the associated Hilbert norm $||.||_{c}=\sqrt{<.\,,.>_{c}}$ (so that $||f||_{c}^{2}=$\\$\int_{0}^{+\infty}|f(t)|^{2}e^{-2ct}dt)$.\\

\hspace*{3mm}Next, we introduce boundary conditions which are naturally motivated by those satisfied by the counting functions of generalized fractal strings.\,Indeed, if $f\in \mathbb{H}_{c}$ and $f$ is absolutely continuous on $\mathbb{R}$ (i.e., $f\in AC(\mathbb{R})$), then $|f(t)|e^{\mp ct}\to 0$ as $t\rightarrow \pm \infty$, respectively; see Remarks \ref{Rk:2.4} and \ref{Rk:2.7}.\footnote{We wish to thank Maxim Kontsevich for having pointed out an inconsistency in the original formulation of the boundary value problem (during a talk on this subjetc given by the second-named author at the IHES), and Pierre Cartier for an ensuing conversation.}\,The asymptotic condition at $+\infty$ implies that (roughly speaking) the functions $f$ satisfying these \emph{boundary conditions} correspond to elements of the space of fractal strings with dimension $D\leq c$.\footnote{Because the domain $D(\partial_{c})$ of the infinitesimal shift $\partial_{c}$ will consist of absolutely continuous functions (see Equation (\ref{Eq:acf})), these are \emph{natural boundary conditions}, in the sense that they are automatically satisfied by any function $f$ in the domain of $\partial_{c}$ or of a function of $\partial_{c}$, such as the spectral operator $\mathfrak{a}_{c}=\zeta(\partial_{c})$.} \,In the context of the counting functions of ordinary fractal strings, this type of asymptotic conditions was studied in detail in \cite{LaPo2}; see also Remarks \ref{Rk:2.4} and \ref{Rk:2.7}.

\begin{remark}\label{Rk:2.7}
Note that in the original multiplicative variable $x=e^{t}$ and for an ordinary fractal string $\mathcal{L}$, it is shown in \emph{$\cite{LaPo2}$} that if $f(e^{t})=N_{\mathcal{L}}(x)$ is of order not exceeding $($respectively, is precisely of the order of$)$ $x^{c}=e^{ct}$ as $x\to +\infty$ $($i.e., as $t\to +\infty$$)$, then $c\leq D$ $($respectively, $c=D$, the Minkowski dimension of $\mathcal{L}$$)$.\,In addition, it follows from \emph{\cite{La2, LaPo2, La-vF3}} that $($with the same notation as above$)$
\begin{equation}
D=\alpha:=\inf\{\gamma \geq 0:\,N_\mathcal{L}(x)=f(e^{t})=O(e^{\gamma t}),\,\mbox{\emph{as}}\,\,t\to +\infty\},
\end{equation}
\text
and hence, $\alpha$ is  the abscissa of convergence of the geometric zeta function $\zeta_{\mathcal{L}}$.\\
\hspace*{3mm}Moreover, let us suppose that $\mathcal{L}$ is normalized so that its geometric counting function satisfies $N_{\mathcal{L}}(x)=0$ for $0<x\leq 1$ $($which, in the additive variable $t=\log x$, amounts to assuming that $f(t)= 0$ for all $t\leq 0$, where $f(t):=N_{\mathcal{L}}(e^{t}))$.\footnote{Without loss of generality, this can always be done since there exists $x_{0}>0$ such that $N_{\mathcal{L}}(x)=0$ for all $0<x<x_{0}$.\,Indeed, it suffices to replace each $l_{j}$ with $\frac{l_{j}}{l_{1}}$ to allow the choice $x_{0}=1$.}\,Then we can simply reflect $f$ with respect to the origin $($i.e., let $F(t):=f(t)$ for $t\geq 0$ and $F(t):=f(-t)$ for $t\leq 0$$)$ in order to obtain an even, nonnegative function $F$ defined on all of $\mathbb{R}$ and therefore having the same asymptotic behavior as $f(|t|)$ as $t\to \pm \infty$.\,In particular, if $f\in L^{2}([0,+\infty), \mu_{c}(dt))$ satisfies $f(t)=o(e^{ct})$ as $t \to +\infty$, then $F\in \mathbb{H}_{c}$ and satisfies the above boundary conditions$:\,F(t)=o(e^{\pm ct})$ as $t \to \pm \infty$, respectively.\,Note that if, furthermore, $f$ is absolutely continuous on $[0,+\infty)$ $($i.e., $f\in AC([0,+\infty))$$)$, then $F$ is absolutely continuous on $\mathbb{R}$ and hence belongs to the domain of $\partial_{c}$, as defined by Equation \emph{(\ref{Eq:acf})} below.\\
\end{remark}

\hspace*{3mm}Now, we precisely define the domain of the differentiation operator $\partial=\partial_{c}$, also called the \emph{infinitesimal shift} (recall the heuristic discussion following Equation (\ref{Eq:EPr})):

\begin{equation}
D(\partial_{c})=\{f\in \mathbb{H}_{c}\cap AC(\mathbb{R}):\,f'\in \mathbb{H}_{c}\},\label{Eq:acf}
\end{equation}
\text
where $AC(\mathbb{R})$  is the space of (locally) absolutely continuous functions on $\mathbb{R}$ and $f'$ denotes the derivative of $f$, viewed either as a function or a distribution.\,Recall that for $f\in AC(\mathbb{R})$, $f'$ exists pointwise almost everywhere and is locally integrable, therefore defining a regular distribution.\,(See, e.g., \cite{Fo}.)\footnote{Note that $D(\partial_{c})$ is the weighted Sobolev space $H^{1}(\mathbb{R},\mu_{c}(dt))$; see, e.g., \cite{Br} for the classic case when $c=0$ and hence this space coincides with the standard Sobolev space $H^{1}(\mathbb{R})$.}\,In addition, we let

\begin{equation}
\partial_{c}:=f',\mbox{\quad for\,} f\in D(\partial_{c}).
\end{equation}

\hspace*{3mm}Our first result will enable us to form various functions of the first order differential operator $\partial_{c}$ and, in particular, to precisely define the spectral operator $\mathfrak{a}_{c}$.

\begin{theorem}\label{Thm:part}
\emph{\cite{HerLa1}}\,$\partial_{c}$ is an unbounded normal \footnote{Recall that this means that $\partial_{c}$ is a closed (and densely defined) operator which commutes with its adjoint $\partial^{*}_{c}$; see \cite{Ru}.} linear operator on $\mathbb{H}_{c}$.\,Moreover, its adjoint $\partial_{c}^{*}$ is given by 

\begin{equation}\label{Eq:part}
\partial_{c}^{*}=2c-\partial_{c},\mbox{\quad\emph{with}\,}\,D(\partial_{c}^{*})=D(\partial_{c}).
\end{equation}
\end{theorem}

\hspace*{3mm}The following restatement of Theorem \ref{Thm:part} may make the situation more transparent to the reader.

\begin{corollary}\label{CorTh}
The normal operator $\partial_{c}$ is given by
\begin{equation}
\partial_{c}=c+iV_{c},\label{Eq:partV}
\end{equation}
where $Re(\partial_{c})=c$, $V_{c}=Im(\partial_{c})$ denote respectively the real and imaginary parts of $\partial_{c}$, and $V_{c}$ is an unbounded self-adjoint operator with the same domain as $\partial_{c}:D(V_{c})=D(\partial_{c})$, as given in Equation \emph{(\ref{Eq:acf})}.\footnote{For notational simplicity, we write $c$ instead of $c$ times the identity operator of $D(\partial_{c})=D(V_{c})$.}
\end{corollary}

\begin{remark}\label{Rk:Vc}
Theorem \ref{Thm:part} $($or equivalently, Corollary \ref{CorTh}$)$ can be proved by showing that $V_{c}=\frac{\partial_{c}-c}{i}$ is unitarily equivalent to $V_{0}$ and using the well-known fact $($see, e.g., \emph{\cite{Sc}} or vol.\,II of \emph{\cite{ReSi}}$)$ according to which the standard momentum operator $V_{0}=\frac{1}{i}\partial_{0}$ is an unbounded self-adjoint operator in $L^{2}(\mathbb{R})$ $($since, via the Fourier transform, it is unitarily equivalent to the multiplication operator by the variable $t$ in $L^{2}(\mathbb{R})=L^{2}(\mathbb{R},dt)=\mathbb{H}_{0}$$)$.\,More specifically, it is shown in \emph{\cite{HerLa1}} that $V_{0}=WV_{c}W^{-1}$ $($or equivalently, $V_{c}=W^{-1}V_{0}W$$)$, where $W:\mathbb{H}_{c}\to \mathbb{H}_{0}$ is the unitary map from $\mathbb{H}_{c}$ onto $\mathbb{H}_{0}$ defined by $(Wf)(t)=e^{-ct}f(t)$ for $f\in\mathbb{H}_{c}$ $($so that $(W^{-1}g)(t)=e^{ct}g(t)$ for $g\in \mathbb{H}_{0}$$)$.
\end{remark}

\begin{remark}\label{Rk:assomom}
It follows from \emph{(\ref{Eq:part})} $($or equivalently, Equation \emph{(\ref{Eq:partV})}$)$ that $\partial^{*}$ is skew-adjoint $($i.e, $\partial^{*}=-\partial_{c}$$)$ or equivalently, that the associated `\emph{momentum operator}'\,$V_{c}=\frac{1}{i} (\partial_{c}-c)$ is self-adjoint if and only if $c=0$.\,Note that this then corresponds to the usual situation of a quantum mechanical particle constrained to move on the real line $\mathbb{R}$.
\end{remark}

In order to find the spectrum $\sigma(\mathfrak{a}_{c})$ of the spectral operator, we first determine the spectrum of $\partial_{c}$, which turns out to be equal to the vertical line of the complex plane passing through the constant $c$.

\begin{theorem}\label{Thm:spp}
\emph{\cite{HerLa1}}\,Let $c\geq 0$.\,Then, the spectrum of $\partial_{c}$ is the closed vertical line of the complex plane passing through $c>0$.\,Furthermore, it coincides with the essential spectrum, $\sigma_{e}(\partial_{c})$, of $\partial_{c}:$
\begin{equation}\label{Eq:spc}
\sigma(\partial_{c})=\sigma_{e}(\partial_{c})=\{\,\lambda\in \mathbb{C}:\,Re(\lambda)=c\,\}.
\end{equation}
In other words, with the notation of Corollary \ref{CorTh}, the spectrum of the self-adjoint operator $V_{c}=Im(\partial_{c})$ is given by
\begin{equation}
\sigma(V_{c})=\sigma_{e}(V_{c})=\mathbb{R}.
\end{equation}
\end{theorem}

\begin{remark}
More specifically, the point spectrum of $\partial_{c}$ is empty $($i.e., $\partial_{c}$ does not have any eigenvalues$)$ and that $\sigma_{ap}(\partial_{c})$, the \emph{approximate point spectrum} of $\partial_{c}$, coincides with $\sigma(\partial_{c})$.\,Hence, $\sigma_{ap}(\partial_{c})$ is also given by the right-hand side of Equation \emph{(\ref{Eq:spc})}.\footnote{Recall that $\lambda \in \sigma_{ap}(\partial_{c})$ (i.e., $\lambda$ is an \emph{approximate eigenvalue} of $\partial_{c}$) if and only if there exists a sequence $\{f_{n}\}^{\infty}_{n=1}$ of elements of $D(\partial_{c})$ such that $||f_{n}||=1$ for all $n\geq 1$ and $||\partial_{c}f_{n}-\lambda f_{n}||_{c}\to 0$ as $n\to \infty$. (See, e.g., \cite{EnNa} or \cite{Sc}.)}\,An entirely analogous statement holds for the spectrum of the `$c$-momentum' operator $V_{c}$ which, according to Remark \ref{Rk:Vc}, coincides with the spectrum of the classic momentum operator $V_{0}$.
\end{remark}

\begin{remark}
For more information about spectral theory, especially in the present case of unbounded operators, we refer to the books \emph{\cite{DunSch, EnNa, Ha, Kat, ReSi, Sc}}.\,The case of unbounded normal operators is treated in \emph{\cite{Ru}}, while that of extended sectorial operators is the object of Haase's monograph \emph{\cite{Ha}}.\,Note that according to Theorem \ref{Thm:part}, the infinitesimal shift $\partial_{c}$ is unbounded $($since, by Theorem \ref{Thm:spp}, its spectrum is unbounded$)$, normal $($by Theorem \ref{Thm:part}, $\partial^{*}_{c}\partial_{c}=\partial_{c}\partial^{*}_{c}$$)$, and sectorial $($in the extended sense of \emph{\cite{Ha}}, since by Theorem \ref{Thm:spp}, $\sigma(\partial_{c})$ is contained in a sector of angle $\frac{\pi}{2}$$)$.\,We caution the reader that the terminology concerning the spectra of unbounded operators is not uniform throughout the well-developed literature on this classical subject.\,For the connections between spectral theory and mathematical physics, mostly in the case of $($unbounded$)$ self-adjoint operators, we refer to \emph{\cite{ReSi}} and \emph{\cite {Sc}}.\,Finally, we note that the present case of normal and $m$-accretive operators\footnote{It is shown in \cite{HerLa1} that in addition to being normal, $\partial_{c}$ is $m$-accretive, in the sense of \cite{Kat}.\,According to a well-known theorem, this means that $\partial_{c}$ is the infinitesimal generator of a contraction semigroup of operators; see Remarks \ref{Rk:2.14} and \ref{Rk:2.15}.} plays an important role in \emph{[JoLa, \S11.6, \S13.5 and \S13.6]} $($in connection with Feynman integrals with singular potentials$)$ and \emph{[La 5, Chapter 5, esp.\,\S5.5]} $($also in connection with the analytic continuation of Feynman integrals with highly singular potentials but in addition, in relation to a conjectural renormalization flow associated with the approach to the Riemann hypothesis developed in \emph{\cite{La5}}; see \emph{[La5, \S5.5.3a and \S5.5.3b]}$)$. 
\end{remark}

\hspace*{3mm}Within our framework in \cite{HerLa1}, we defined the \textit{spectral operator} as follows:
\begin{equation}
\mathfrak{a}=\zeta(\partial),\mbox{\quad where $\partial=\partial_{c}$ and $\mathfrak{a}=\mathfrak{a}_{c}$,}
\end{equation}
via the measurable functional calculus for unbounded normal operators; see, e.g., \cite{Ru}.\,(If, for simplicity, we assume $c\ne 1$ to avoid the pole of $\zeta$ at $s=1$, then $\zeta$ is holomorphic (and, in particular, continuous) in an open neighborhood of $\sigma(\partial)$.\,If $c=1$ is allowed, then $\zeta$ is meromorphic in an open neighborhood of $\sigma(\partial)$ (actually, in all of $\mathbb{C}$).\,Hence, when $c\ne 1$, we could simply use the holomorphic (or the continuous) functional calculus for unbounded normal operators (see \cite{Ru}), whereas when $c=1$, we could use the meromorphic functional calculus for sectorial operators\,(see \cite{Ha}).\\

\hspace*{3mm}The domain of the spectral operator is the following:

\begin{equation}\label{Eq:Dsp}
D(\mathfrak{a})=\{f\in D(\partial):\,\mathfrak{a}(f)=\zeta(\partial)(f)\in \mathbb{H}_{c}\}.
\end{equation}

\hspace*{3mm}Our next result, Theorem \ref{Thm:11} below, provides a representation of the spectral operator $\mathfrak{a}$ as a composite map of the Riemann zeta function $\zeta$ and the first order differential operator $\partial_{c}$.\,It also gives a natural connection between this representation and the earlier one obtained for the spectral operator in Equations (\ref{Eq:spop}) and (\ref{Eq:Spr}).\,(See also Remarks \ref{Rk:2.14}, \ref{Rk:2.15} and \ref{Rk:Ep} below.)

\begin{theorem}\emph{\cite{HerLa1}}\,Assume that $c>1$.\,Then, $\mathfrak{a}$ can be uniquely extended to a bounded operator on $\mathbb{H}_{c}$ and, for any $f\in \mathbb{H}_{c}$, we have $($for almost all $t\in \mathbb{R}$ or as an equality in $\mathbb{H}_{c}$$):$\label{Thm:11}
\begin{equation}
\mathfrak{a}(f)(t)=\sum_{k=1}^{\infty}f(t-\log k)=\zeta(\partial_{c})(f)(t)=\left(\sum_{n=1}^{\infty}n^{-\partial_{c}}\right)(f)(t).\label{Eq:spcf}
\end{equation}
\end{theorem}
In other words, for $c>1$, we have

\begin{equation}
\mathfrak{a}_{c}=\zeta(\partial_{c})=\sum_{n=1}^{\infty}n^{-\partial_{c}},\label{Eq:2.23}
\end{equation}
\text
where the equality holds in $\mathcal{B}(\mathbb{H}_{c})$, the space of bounded linear operators on $\mathbb{H}_{c}$.\\

\hspace*{3mm}For any $c>0$, we also show in \cite{HerLa1} that Equation (\ref{Eq:spcf}) holds for all $f$ in a suitable dense subspace of $D(\mathfrak{a})$, which we conjectured to be a core for $\mathfrak{a}$ and hence to uniquely determine the unbounded operator $\mathfrak{a}=\mathfrak{a}_{c}=\zeta(\partial_{c})$, viewed as the `\emph{analytic continuation}' of $\sum_{n=1}^{\infty}n^{-\partial_{c}}$ to the critical strip.\\

\begin{remark}\label{Rk:2.14}
Note that the strongly continuous contraction group of \\bounded linear operators $\{e^{-t\partial_{c}}\}_{t\in \mathbb{R}}$ plays a crucial role in the earlier representation obtained in Theorem \ref{Thm:11} of the spectral operator $\mathfrak{a}_{c}=\zeta(\partial_{c})$.\,We refer to \emph{\cite{EnNa, Go, HiPh, Kat}} for the theory of strongly continuous semigroups.
\end{remark}

\begin{remark}\label{Rk:2.15}
Using Corollary \ref{CorTh}, we show in \emph{\cite{HerLa1}} that $\{e^{-t\partial_{c}}\}_{t\in \mathbb{R}}$ is a strongly continuous contraction group of operators and $||e^{-t\partial_{c}}||= e^{-tc}$ for any $t\in\mathbb{R}$ and any $c\geq0$.\,The adjoint group $\{(e^{-t\partial_{c}})^{*}\}_{t\in \mathbb{R}}$ is then given by $\{e^{-t\partial_{c}^{*}}\}_{t\in\mathbb{R}}=\{e^{-t(2c-\partial_{c})}\}_{t\in\mathbb{R}}$.\,The strongly continuous group of operators $\{e^{-t\partial_{c}}\}_{t\in \mathbb{R}}$ is a translation $($or \emph{shift}$)$ group.\,That is, for every $t\in \mathbb{R}$, $(e^{-t\partial_{c}})(f)(u)=f(u-t)$, for all $f\in \mathbb{H}_{c}$ and $u\in \mathbb{R}$.\,$($For a fixed $t\in \mathbb{R}$, this equality holds between elements of $\mathbb{H}_{c}$ and hence, for a.e. $u\in\mathbb{R}$$.)$\,As a result, the infinitesimal generator $\partial=\partial_{c}$ of the shift group $\{e^{-t\partial}\}_{t\in \mathbb{R}}$ is called the \emph{infinitesimal shift} of the real line. 
\end{remark}

\hspace*{3mm}In order to study the invertibility of the spectral operator, it is necessary to determine the spectrum of $\mathfrak{a}_{c}$ (denoted by $\sigma(\mathfrak{a}_{c})$).\,Using the Spectral Mapping Theorem for unbounded normal operators (the continuous version when $c\ne 1$ and the meromorphic version, when $c=1$, see the appropriate appendix in \cite{HerLa1}), we obtain the following characterization of $\sigma(\mathfrak{a}_{c})=\sigma(\mathfrak{a})$:\footnote{We wish to acknowledge helpful written exchanges with Rainer Nagel, Tanja Eisner, Markus Haase and Daniel Lenz, about appropriate versions of the Spectral Mapping Theorem in this context.}

\begin{theorem}\emph{\cite{HerLa1}}\,Assume that $c\geq0$.\,Then \label{Thm:ssop}
\begin{equation}\label{Eq:Merv}
\sigma(\mathfrak{a})=\overline{\zeta(\sigma(\partial))}=cl\big(\zeta(\{\lambda\in\mathbb{C}:\,Re(\lambda)=c\})\big),
\end{equation}
where $\sigma(\mathfrak{a})$\,is the spectrum of $\mathfrak{a}=\mathfrak{a}_{c}$ and $\overline{N}=cl(N)$ is the closure of $N\subset \mathbb{C}$.
\end{theorem}

\begin{center}
\section{Invertibility of the Spectral Operator $\mathfrak{a}_{c}$} 
\end{center}

To study the invertibility of the spectral operator $\mathfrak{a}_{c}$, we first introduce two notions of invertibility, namely \emph{quasi-invertibility} and \emph{almost invertibility} (see Definitions \ref{Def:qi} and \ref{Def:ai} below).\,We show that these two notions of invertibility play an important role in unraveling the precise relation between the existence of a suitable `inverse' for the spectral operator and the inverse spectral problem for fractal strings .\\

\hspace*{3mm}Our next result shows that the quasi-invertibility of $\mathfrak{a}_{c}$ is intimately connected with the location of the critical zeroes of the Riemann zeta function and hence, with the Riemann hypothesis.\\

\begin{theorem}\emph{\cite{HerLa1}}\,Assume that $c\geq 0$.\,Then, the spectral operator $\mathfrak{a_{c}}=\zeta(\partial_{c})$ is quasi-invertible if and only if the Riemann zeta function does not vanish on the vertical line $\{s\in \mathbb{C}:\,Re(s)=c\}$.
\end{theorem}

\begin{corollary}\emph{\cite{HerLa1}}\,The spectral operator $\mathfrak{a}$ is quasi-invertible for all $c\in(0,1)-\frac{1}{2}$ $($or equivalently, for all $c\in (0,\frac{1}{2})$$)$ if and only if the Riemann hypothesis is true.
\end{corollary}

In order to define the notion of quasi-invertibility, we first introduce the \emph{truncated infinitesimal shift} $\partial^{(T)}$ and the \emph{truncated spectral operator} $\mathfrak{a}^{(T)}$.\,As was recalled in Corollary \ref{CorTh} and Theorem \ref{Thm:spp}, the infinitesimal shift $\partial=\partial_{c}$ is given by
\begin{equation}
\partial_{c}=c+iV,
\end{equation}
\text
where $V=V_{c}$ is an unbounded self-adjoint operator with spectrum $\sigma(V)=\mathbb{R}$.\,Thus, given $T\geq 0$, we define the \emph{truncated infinitesimal shift} as follows:\\
\begin{equation}
A^{(T)}=\partial^{(T)}:=c+iV^{(T)},
\end{equation}
where
\begin{equation}
V^{(T)}:=\phi^{(T)}(V)\notag\\
\end{equation}
\text
and $\phi^{(T)}$ is a suitable (i.e., $T$-admissible) cut-off function (so that $\sigma(A^{(T)})$\\$=c+i[-T,T]$).\footnote{More precisely, $\phi^{(T)}$ is any $T$-\emph{admissible cut-off function}, defined as follows: when $c\ne 1$, $\phi^{(T)}$ is a continous function defined on $\mathbb{R}$ and the closure of its range is equal to $[-T,T]$.\,Furthermore, when $c=1$, $\phi^{(T)}$ is meromorphic in an open neighborhood of $\mathbb{R}$ in $\mathbb{C}$ and the closure of its restriction to $\mathbb{R}$ has range $[-T,T]$; in this case, one views $\phi^{(T)}$ as a continuous function with values in the Riemann sphere $\widetilde{\mathbb{C}}:=\mathbb{C}\cup\{\infty\}$.\,(For example, we may take
$\phi^{(T)}(s)=\frac{T}{\pi}\tan^{-1}(s)$, for $s\in \mathbb{R}$.)\,One then uses the measurable functional calculus for unbounded normal operators, along with the corresponding continuous ($c\ne 1$) or meromorphic ($c=1$) Spectral Mapping Theorem (see the relevant appendix in \cite{HerLa1}) in order to define both $\partial^{(T)}$ and $\mathfrak{a}^{(T)}$ and calculate their spectra.\\}
Similarly, the \emph{truncated spectral operator} is defined by

\begin{equation}
\mathfrak{a}^{(T)}:=\zeta\left(\partial^{(T)}\right).
\end{equation}

\hspace*{3mm}Note that the above construction can be generalized as follows: \\

Given $0\leq T_{0}\leq T$, one can define a $(T_{0},T)$-admissible cut-off function $\phi^{(T_{0},T)}$ exactly as above, except with $[-T,T]$ replaced with $\{\tau\in\mathbb{R}:\,T_{0}\leq |\tau| \leq T\}$.\\

Correspondingly, one can define $V^{(T_{0},T)}=\phi^{(T_{0},T)}$,\\
\begin{equation}
A^{(T_{0},T)}=\partial^{(T_{0},T)}:=c+iV^{(T_{0},T)}
\end{equation}
and
\begin{equation}
\mathfrak{a}^{(T_{0},T)}=\zeta(\partial^{(T_{0},T)}),
\end{equation}
where $\partial^{(T_{0},T)}$ is the $(T_{0},T)$-infinitesimal shift and $\mathfrak{a}^{(T_{0},T)}$ is the $(T_{0},T)$-truncated spectral operator.\,Note that for $T_{0}=0$, we recover $A^{(T)}$ and $\mathfrak{a}^{(T)}$.\\

\hspace*{3mm}Finally, we introduce the notions of \emph{quasi-invertibility}  and \emph{almost invertibility} of $\mathfrak{a}$ as follows:

\begin{definition}\label{Def:qi}
The spectral operator $\mathfrak{a}$ is \emph{quasi-invertible} if its truncation $\mathfrak{a}^{(T)}$ is invertible for all $T>0$.
\end{definition}

\begin{definition}\label{Def:ai}
Similarly, $\mathfrak{a}$ is \emph{almost invertible} if for some $T_{0}\geq T$, its truncation $\mathfrak{a}^{(T_{0},T)}$ is invertible for all $T>0$.
\end{definition}

Note that in the definition of \textquotedblleft almost invertibility\textquotedblright, $T_{0}$ is allowed to depend on the parameter $c$.\,Furthermore, observe that quasi-invertiblity implies almost invertibility.\\

\hspace*{3mm}The spectra of $A^{(T)}$ and $\mathfrak{a}^{T}$ are now determined as follows:

\begin{theorem}\emph{\cite{HerLa1}}\,For all $T>0$, $A^{(T)}$ and $\mathfrak{a}^{(T)}$ are bounded normal operators,\footnote{More precisely, only when $c=1$, which corresponds to the pole of $\zeta(s)$ at $s=1$, $\mathfrak{a}^{(T)}$ is not bounded (since $\zeta(1)=\infty\in \widetilde{\mathbb{C}}$) and hence, Equation (\ref{Eq:trsp}) must then be interpreted as an equality in $\widetilde{\mathbb{C}}$.} with spectra respectively given by \label{Thm:spcqt}
\begin{equation}
\sigma(A^{(T)})=\{c+i\tau:\,|\tau|\leq T\},\label{Eq:trpr}
\end{equation}

\begin{equation}
\sigma(\mathfrak{a}^{(T)})=\{\zeta(c+i\tau):\,|\tau|\leq T \}.\label{Eq:trsp}
\end{equation}
\end{theorem}

Recall that by definition of the spectrum, $\mathfrak{a}^{(T)}$ is invertible if and only if $0\notin \sigma(\mathfrak{a}^{(T)})$.\\

\hspace*{3mm}More generally, given $0\leq T_{0}\leq T$, the exact counterpart of Theorem \ref{Thm:spcqt} holds for $A^{(T_{0},T)}$ and $\mathfrak{a}^{(T_{0},T)}$, except with $|\tau|\leq T$ replaced with $T_{0}\leq |\tau|\leq T$.\, A similar comment can be made about Corollary \ref{Cor:trun} below.\\

\hspace*{3mm}Our next result provides a necessary and sufficient condition for the invertibility of the truncated spectral operator.\\

\begin{corollary}\emph{\cite{HerLa1}}\,Assume that $c\geq 0$.\,Then, the truncated spectral operator $\mathfrak{a}^{(T)}$ is invertible if and only if $\zeta$ does not vanish on the vertical line segment $\{s\in \mathbb{C}:\,Re(s)=c,\,|Im(s)|\leq T\}$.\label{Cor:trun}
\end{corollary}

Naturally, given $0\leq T_{0}\leq T$, the same result is true for $\mathfrak{a}^{(T_{0},T)}$ provided $|Im(s)|\leq T$ is replaced with $T_{0}\leq |Im(s)|\leq T$.\\

Next, we deduce from the above results necessary and sufficient conditions ensuring the quasi-invertibility or the almost invertibility of $\mathfrak{a}_{c}$.\,Such conditions turn out to be directly related to the location of the critical zeroes of the Riemann zeta function.\footnote{In light of Definitions \ref{Def:qi} and \ref{Def:ai} , Theorem \ref{Thm:qual} follows from Corollary \ref{Cor:trun}.}

\begin{theorem}\emph{\cite{HerLa1}}\,Assume that $c\geq 0$.\,Then,\label{Thm:qual}

\begin{enumerate}
\item $\mathfrak{a}$ is quasi-invertible if and only if $\zeta$ does not vanish $($i.e., does not have any zeroes$)$ on the vertical line $Re(s)=c$.
\item $\mathfrak{a}$ is almost invertible if and only if all but $($at most$)$ finitely many zeroes of $\zeta$ are off the vertical line $Re(s)=c$.
\end{enumerate}
\end{theorem}

We then deduce from Theorem \ref{Thm:qual} a \emph{spectral reformulation of the Riemann hypothesis} (see Corollary \ref{Cor:RHy} below).\,These results recapture from a rigorous operator theoretic point of view, and further extend, the work of the second author and H.\,Maier in their study of \emph{the inverse spectral problem for vibrating fractal strings},\,which answers the question
 
\begin{quotation}
\begin{center}
\textquotedblleft \emph{Can one hear the shape of a fractal string}?\textquotedblright 
\end{center}
\end{quotation}
or equivalently, 

\begin{quotation}
\textquotedblleft \emph{To what extent could the geometry of a fractal string be recovered from its spectrum}?\textquotedblright\\
\end{quotation}
(See \cite{LaMa1,LaMa2}; see also the beginning of \S2 above and [La-vF3, Chapter 9].)\,They also give a precise mathematical meaning in this context to the notion of invertibility of the spectral operator, as discussed semi-heuristically in [La-vF3, Corollary 9.6].\\

\hspace*{3mm}We note, in particular, that the second part of Theorem \ref{Thm:qual} above and of Corollary \ref{Cor:RHy} just below does not have any counterpart in \cite{LaMa1, La-vF3} or, to our knowledge, in the extensive literature aimed at providing various reformulations of the Riemann hypothesis.

\begin{remark}
The criteria provided in Theorem \ref{Thm:qual} and Corollary \ref{Cor:RHy} clearly extend in a natural manner to the spectral operators associated to a large class of  arithmetic zeta functions $($or L-functions$)$.\,The same can be said of most of the results of \emph{\cite{HerLa1, HerLa2}} discussed in this survey.
\end{remark}

\hspace*{3mm}As a result of Theorem \ref{Thm:qual}, the invertibility of the spectral operator $\mathfrak{a_{c}}$ can be characterized as follows:\footnote{In light of the functional equation satisfied by $\zeta$, we could equivalently replace the interval $(\frac{1}{2},1)$ with $(0,\frac{1}{2})$ and the inequality $c>\frac{1}{2}$ with $c<\frac{1}{2}$ in the statement of Corollary \ref{Cor:RHy}.}

\begin{corollary}\emph{\cite{HerLa1}}\,Assume that $c\geq 0$.\,Then,\label{Cor:RHy}

\begin{enumerate}
\item $\mathfrak{a}$ is quasi-invertible for all $c\in (\frac{1}{2},1)$ if and only if the Riemann hypothesis is true.
\item $\mathfrak{a}$ is almost invertible for all $c\in (\frac{1}{2},1)$ if and only if the Riemann hypothesis (RH) is \textquotedblleft almost true \textquotedblright $($i.e., on every vertical line $Re(s)=c$, $c>\frac{1}{2}$, there are at most finitely many exceptions to RH$)$.
\end{enumerate}

\end{corollary}

\begin{remark}
Note that, according to our previous results and definitions, the invertibility of the spectral operator $\mathfrak{a}$ implies its quasi-invertibility, which also implies its almost invertibility.
\end{remark}

\begin{corollary}\emph{\cite{HerLa1}}\,For $c=\frac{1}{2}$, the spectral operator $\mathfrak{a}$ is not almost $($and thus, not quasi-$)$ invertible.\,$($This follows from Hardy's theorem according to which $\zeta$ has infinitely many zeroes on the critical line $Re(s)=\frac{1}{2}$.$)$ 
\end{corollary}

\hspace*{3mm}Recall that saying that $\mathfrak{a}$ is \emph{invertible} means that $\mathfrak{a}$ is a bijection from its domain $D(\mathfrak{a})$ onto $\mathbb{H}_{c}$ and that its inverse $\mathfrak{a}^{-1}$ is a bounded linear operator on $\mathbb{H}_{c}$ (i.e., $\mathfrak{a}^{-1}\in \mathcal{B}(\mathbb{H}_{c})$).\,Since $\mathfrak{a}$ is a closed operator (because it is normal), this is equivalent to the bijectivity of $\mathfrak{a}$ from $D(\mathfrak{a})$ onto $\mathbb{H}_{c}$ (in light of the Closed Graph Theorem, see \cite{Fo} or \cite{Ru}).\\

\hspace*{3mm}Moreover, by definition of the spectrum $\sigma(\mathfrak{a})$ of $\mathfrak{a}$, the operator $\mathfrak{a}$ is invertible if and only if $0\notin \sigma(\mathfrak{a})$.\,We therefore deduce from Theorem \ref{Thm:ssop} (the characterization of the spectrum of $\mathfrak{a}$) the following \emph{invertibility criterion} for $\mathfrak{a}$.

\begin{theorem}\emph{\cite{HerLa1}}\,Assume that $c\geq 0$.\,Then, the spectral operator $\mathfrak{a}$ is invertible if and only if $0\notin cl(\{\zeta(s):\,Re(s)=c\})$.\footnote{That is, if and only if $\zeta$ does not have any zeroes on the vertical line $L_{c}:=\{Re(s)=c\}$ and there is no infinite sequence $\{s_{n}\}_{n=1}^{\infty}$ of distinct points of $L_{c}$ such that $\zeta(s_{n})\to 0$ as $n\to \infty$.}\label{Thm:inv}
\end{theorem}

\hspace*{3mm}We will explore the consequences of Theorem \ref{Thm:inv} in \S4, in light of the universality of $\zeta(s)$ in the right critical strip $\{\frac{1}{2}<Re(s)<1\}$ and conditionally) of the non-universality of $\zeta(s)$ in the left critical strip $\{0<Re(s)<\frac{1}{2}\}$.\,For now, we limit ourselves to examining the case when $c>1$.\,[Theorem \ref{Thm:quasinv} just below can be viewed as a corollary of our previous results (when combined with some analytic estimates of $\zeta(s)$ for $Re(s)>1$), and especially of Theorem \ref{Thm:qual}.\,However, when combined with the result of \cite{HerLa2} stated in Remark \ref{Rk:Ep} below, it provides more specific information about $\mathfrak{a}$ (and its inverse $\mathfrak{a}^{-1}$).]

\begin{theorem}\emph{\cite{HerLa1}}\,For $c>1$, $\mathfrak{a}$ is quasi- $($and hence, almost$)$ invertible.\,In fact, we will see next that $\mathfrak{a}$ is also invertible for $c>1$.\label{Thm:quasinv}
\end{theorem}

\begin{remark}\label{Rk:Ep}One can show $($see \emph{\cite{HerLa2}}$)$ that for $c>1$, $\mathfrak{a}$ belongs to $\mathcal{B}(\mathbb{H}_{c})$ and is given by the following \emph{operator-valued Euler product expansion:}

\begin{equation}
\mathfrak{a}=\zeta(\partial)=\prod_{p\in\mathcal{P}}(1-p^{-\partial})^{-1},\label{Eq:EPrd}
\end{equation}
\text
where the convergence holds in the space  $\mathcal{B}(\mathbb{H}_{c})$ of bounded linear operators on $\mathbb{H}_{c}$.\,Furthermore,

\begin{equation}
||\mathfrak{a}||\leq \zeta(c)
\end{equation}
\text
and $($still for $c>1$$)$ $\mathfrak{a}$ is invertible with $($bounded$)$ inverse given by

\begin{equation}
\mathfrak{a}^{-1}=\frac{1}{\zeta}(\partial)=\sum_{n=1}^{\infty}\mu(n)n^{-\partial},\label{Eq:3.10}
\end{equation}
\text
where the equality holds in $\mathcal{B}(\mathbb{H}_{c})$ and $\mu(n)$ is the M\"obius function defined by $\mu(n)=(-1)^{q}$ if $n\in \mathbb{N}$ is a product of $q$ distinct primes, and $\mu(n)=0$, otherwise.\,$($Compare Equations \emph{(\ref{Eq:3.10})} and \emph{(\ref{Eq:2.23})}.$)$\\
 
\hspace*{3mm}Moreover, it was conjectured in \emph{[La-vF3, \S6.3.2]} that the above Euler product \emph{(\ref{Eq:EPrd})} also converges $($in a suitable sense$)$ inside the critical strip, that is, for $0<c<1$.\,This conjecture is addressed in \emph{\cite{HerLa2}}.
\end{remark}

\begin{center}
\section{Phase Transitions} 
\end{center}

The connection between the values of the Riemann zeta function and the spectrum of $\mathfrak{a}_{c}$ enabled us to observe, within our framework, the existence of \emph{phase transitions} at $c=\frac{1}{2}$ and $c=1$ for the spectral operator $\mathfrak{a}_{c}=\zeta(\partial_{c})$.\,These mathematical phase transitions are related to the quasi-invertibility of $\mathfrak{a}_{c}$, the boundedness and unboundedness of $\mathfrak{a}_{c}$ (or equivalently, of its spectrum $\sigma(\mathfrak{a}_{c})$), as well as to the universality and invertibility of $\mathfrak{a}_{c}$.\\

\subsection{Phase transitions for the boundedness and invertibility of $\mathfrak{a}_{c}$}

\textbf{I.\,Phase Transition at $c=\frac{1}{2}$:}

\vspace*{2mm}
\begin{theorem}\emph{\cite{HerLa1}}\,Conditionally $($i.e., under the Riemann hypothesis$)$,\\the spectral operator $\mathfrak{a}_{c}$ is quasi-invertible for all $c\ne\frac{1}{2}$ $($$c\in (0,1)$$)$, and $($unconditionally$)$ it is not quasi-invertible $($not even almost invertible$)$ for $c=\frac{1}{2}$.\label{Thm:q-v}
\end{theorem}

\begin{remark}
Recall that assuming that $\mathfrak{a}_{c}$ is quasi-invertible for all $c\ne \frac{1}{2}$ is equivalent to the truth of the Riemann hypothesis.\,Furthermore, $\mathfrak{a}_{c}$ is not almost invertible for $c=\frac{1}{2}$ because $\zeta$ has infinitely many zeroes on the critical line $Re(s)=\frac{1}{2}$.
\end{remark}

\textbf{II.\,Phase Transitions at $c=\frac{1}{2}$ and $c=1$:}

\vspace*{2mm}
\begin{theorem}\emph{\cite{HerLa1}}\,The spectral operator $\mathfrak{a}_{c}$ is invertible for $c>1$, is not invertible for $\frac{1}{2}<c<1$ $($conjecturally, also for $c=\frac{1}{2}$$)$, and invertible $($conditionally$)$ for $0<c<\frac{1}{2}$.\label{Thm:ci}
\end{theorem}

\vspace*{2mm}
\begin{theorem}\emph{\cite{HerLa1}}\,The spectrum $\sigma(\mathfrak{a}_{c})$ is non-compact $($and hence, unbounded$)$, but $($conditionally$)$ not all of $\mathbb{C}$, for $0<c<\frac{1}{2}$.\,It is all of $\mathbb{C}$ for $\frac{1}{2}<c<1$, and compact $($and thus, bounded$)$ for $c>1$.\label{Thm:sun}
\end{theorem}

\textbf{III.\,Phase Transition at $c=1$:}

\vspace*{2mm}
\begin{theorem}\emph{\cite{HerLa1}}\,Unconditionally, the spectral operator $\mathfrak{a}_{c}$ is unbounded for $c\leq1$, and is bounded for $c>1$.\label{Thm:unc}
\end{theorem}

\subsection{Phase transitions for the shape of the spectrum of $\mathfrak{a}_{c}$}

\hspace*{3mm}Recall from Theorem \ref{Thm:ssop} that $\sigma(\mathfrak{a})$ is equal to the closure of the range of $\zeta$ on the vertical line $\sigma(\partial)=\{Re(s)=c\}$.\,Hence, $\sigma(\mathfrak{a})$ is equal to $\{\zeta(c+i\tau):\,\tau\in\mathbb{R}\}$ union its limit points in $\mathbb{C}$.\,Also, by definition of $\sigma(\mathfrak{a})$,\,$\mathfrak{a}$ is invertible if and only if $0\notin \sigma(\mathfrak{a})$.\,As a consequence, we obtain the following result, which synthetizes and complements Theorems \ref{Thm:ci}, \ref{Thm:sun} and \ref{Thm:unc} about phase transitions for $\mathfrak{a}_{c}$ at both $c=1$ and $c=\frac{1}{2}$.\\

\begin{theorem}\emph{\cite{HerLa1}}\,Assume that $c\geq 0$.\,Then,\label{ThM}
\begin{enumerate}
\item For $c>1$, $\sigma(\mathfrak{a})$ is bounded and $0\notin \sigma(\mathfrak{a})$.\,Hence, $\mathfrak{a}$ is invertible.\\Furthermore, $\sigma(\mathfrak{a})$ is unbounded for $c\leq 1$.
\item$($Universality$)$ For $c\in (\frac{1}{2},1)$, $\sigma(\mathfrak{a})=\mathbb{C}$.\,$($This follows from the Bohr--Landau Theorem and, more generally, from the universality of $\zeta$ in the right critical strip $\frac{1}{2}<Re(s)<1$.$)$\,Hence, $\mathfrak{a}$ is \emph{not} invertible.
\item For $c\in (0,\frac{1}{2})$, $\sigma(\mathfrak{a})$ is unbounded and conditionally $($i.e., under RH$)$, $\sigma(\mathfrak{a})\ne \mathbb{C}$ and $\mathfrak{a}$ is invertible $($because $0\notin \sigma(\mathfrak{a})$$)$.
\end{enumerate}
\end{theorem}

Note that the last statement in the third part of the theorem follows from the non-universality of $\zeta$ in the left critical strip $0<Re(s)<\frac{1}{2}$; see \cite{GarSt}.\\

\begin{remark}
We should stress that we are talking here about phase transitions in \emph{the mathematical sense} and \emph{not} in the physical sense commonly used in condensed matter physics, statistical physics and the theory of critical phenomena.\,However, there are some intriguing analogies between the two types of transitions.\,Namely, in the present situation, there is an abrupt change in the mathematical properties of the system at certain critical values of the parameter $c$, here at $c=1$ and more important, at $c=\frac{1}{2}$.
\end{remark}

\begin{remark}
In \emph{\cite{HerLa1}}, we also obtain an operator theoretic counterpart of Voronin's Theorem and its later extensions according to which the Riemann zeta function $\zeta=\zeta(s)$ is universal in the right critical strip $\frac{1}{2}<Re(s)<1$.\,\textquotedblleft Universality\textquotedblright \,in this context means that any non-vanishing holomorphic function $($in a suitable subregion$)$ can be uniformly approximated by imaginary translates of $\zeta$; see, e.g., the books \emph{\cite{KarVo, Lau, St}}.\,Accordingly, in \emph{\cite{HerLa1}}, we show that the spectral operator $\mathfrak{a}=\zeta(\partial)$ is universal $($in an appropriate sense$)$ among a large class of operators obtained via the functional calculus for the infinitesimal shift $\partial$.\footnote{More specifically, they are obtained via the holomorphic (or continuous) functional calculus for the truncated spectral operators $\partial^{(T)}=\partial_{c}^{(T)}$.}\,We note that the universality of $\zeta$ was used in several of the theorems of \emph{\cite{HerLa1}} discussed in this section.
\end{remark}

\begin{center}
\section{Concluding Comments}
\end{center}

In \cite{La3}, the second author pointed out that it would be interesting to interpret the reformulation of the Riemann hypothesis obtained in [LaMa1, 2] as a mathematical phase transition at the `\emph{critical fractal dimension}' $D=\frac{1}{2}$ (corresponding in our present situation to the critical parameter $c=\frac{1}{2}$).\,From a broader prospective, he went on to ask whether one could reinterpret the then nascent mathematical theory of complex fractal exponents and dimensions in terms of Wilson's renormalization theory  aimed at understanding physical phase transitions and critical phenomena via the analytic continuation (in a complexified space-time parameter) of certain Feynman-type integrals; see, especially, [La3, Question 2.6, p.147].\,This line of inquiry was pursued by the second author in aspects of [La5, esp.\,Chapter 5]; see, in particular, [La5, \S5.5 and \S5.6].\,Along related lines, it would be interesting to investigate the possible relationships between the mathematical phase transition occurring in our work at the critical dimension $D=1$ (i.e., $c=1$) and the phenomenon of symmetry breaking established by Jean-Benoit Bost and Alain Connes [BosCon1,2] in connection with the pole at $s=1$ of the Riemann zeta function, using tools from operator algebras (especially, KMS states and the Tomita--Takesaki theory) as applied to quantum statistical mechanics.\\

\hspace*{3mm}From our present perspective, however, the phase transition observed in the mid-fractal case when $c=\frac{1}{2}$ is the most important and intruiging one, since it correspond to the critical line $Re(s)=\frac{1}{2}$ and has multiple origins and interpretations (namely, a reformulation of the Riemann hypothesis as in [LaMa1,2], in [La-vF2,3] and as in Theorem \ref{Thm:q-v} above, transition between non-universality and universality of the Riemann zeta function $\zeta=\zeta(s)$, and hence, as in Theorems \ref{Thm:ci} and \ref{ThM} above, between invertibility and non-invertibility of the spectral operator $\mathfrak{a}=\mathfrak{a}_{c}=\zeta(\partial_{c})$).\,We close this comment by asking whether the mathematical phase transition occurring at the \emph{critical fractal dimension} $c=\frac{1}{2}$ (as evidenced by Theorems \ref{Thm:q-v}, \ref{Thm:sun} and \ref{ThM} above) could also be interpreted physically as a symmetry breaking phenomenon in terms of KMS states for a suitable quantum statistical model (perhaps for an appropriate extension or reinterpretation of the model of \cite{BosCon2} and its sequels).\\

\vspace*{0.5cm}
\textbf{Acknowledgements}:\,The research of M.\,L.\,Lapidus was partially supported by the National Science Fundation under the research grants DMS-0707524 and DMS-1107750.\\

\vspace*{0.5cm}
 \begin{center}
\begin{tabular*}{0.96\textwidth}{l@{\extracolsep{\fill}}l}
Hafedh Herichi\\
University of California, Riverside\\
Department of Mathematics\\
Riverside, CA 92521-0135\\
USA\\
\vspace*{1mm}
\texttt{herichi@math.ucr.edu}\\
\end{tabular*}
\end{center}

\begin{center}
\begin{tabular*}{0.96\textwidth}{l@{\extracolsep{\fill}}l}
Michel L.\,Lapidus\\
University of California, Riverside\\
Department of Mathematics\\
Riverside, CA 92521-0135\\
USA\\
\vspace*{1mm}
\texttt{lapidus@math.ucr.edu}\\
\end{tabular*}
\end{center}
   

\begin{thebibliography}{99}

\bibitem[Berr1]{Berr1}
M. V. Berry,
Distribution of modes in fractal resonators, in:
\emph{Structural Stability in Physics} (W. G\"uttinger and H. Eikemeier, eds.), Springer-Verlag, Berlin, 1979, pp. 51--53.

\bibitem[Berr2]{Berr2}
M. V. Berry, 
Some geometric aspects of wave motion:
Wave-front dislocations, diffraction catastrophes, diffractals, in:
\emph{Geometry of the Laplace Operator}, Proc. Sympos. Pure Math., vol. 36, Amer. Math. Soc., Providence, R. I., 1980, pp. 13--38.

\bibitem[BosCon1]{BosCon1}
J.-B. Bost and A. Connes,
Produit eul\'erien et facteurs de type III, \emph{C. R. Acad. Sci. Paris S\'er. I Math}. \textbf{315} (1992), 279--284.


\bibitem[BosCon2]{BosCon2}
J.-B. Bost and A. Connes,
Hecke algebras, type III factors and phase transitions with spontaneous symmetry breaking in number theory, \emph{Selecta Math}. (N.S.) \textbf{1} (1995), 411--457.

\bibitem[Br]{Br}
H. Brezis,
\emph{Functional Analysis, Sobolev Spaces and Partial Differential Equations},
Universitext, Springer, New York, 2011. (English transl. and rev. and enl. ed. of H. Brezis, \emph{Analyse Fonctionelle: Th\'eorie et applications}, Masson, Paris, 1983.)

\bibitem[BroCa]{BroCa}
J. Brossard and R. Carmona, Can one hear the dimension of a fractal?, \emph{Commun. Math. Phys.} \textbf{104} (1986), 103--122.

\bibitem[Coh]{Coh}
D. L. Cohn,
\emph{Measure Theory}, Birkh\"auser, Boston, 1980.

\bibitem[DerGrVo]{DerGrVo}
G. Derfel, P. Grabner and F. Vogl,
The zeta function of the Laplacian on certain fractals, \emph{Trans. Amer. Math. Soc.} \textbf{360} (2008), 881--897.


\bibitem[DunSch]{DunSch}
N. Dunford and J. T. Schwartz,
\emph{Linear Operators}, Parts I-III, Wiley Classics Library, John Wiley \& Sons, Hoboken, 1988. (Part I: \emph{General Theory.} Part II: \emph{Spectral Theory}. Part III: \emph{Spectral Operators}.)


\bibitem[Edw]{Edw}
H. M. Edwards,
\emph{Riemann's Zeta Function},
Academic Press, New York, 1974.

\bibitem[EnNa]{EnNa}
K.-J. Engel and R. Nagel,
\emph{One-Parameter Semigroups for Linear Evolution Equations}, 
Graduate Texts in Mathematics, Springer, Berlin, 2000.  

\bibitem[FlVa]{FlVa}
J. Fleckinger and D. Vassiliev,
An example of a two-term asymptotics for the \textquotedblleft counting function\textquotedblright of a fractal drum, \emph{Trans. Amer. Math. Soc.} \textbf{337} (1993), 99-116.

\bibitem[Fo]{Fo}
G. B. Folland,
\emph{Real Analysis}:
\emph{Modern Techniques and Their Applications}, 2nd ed., John Wiley \& Sons, Boston, 1999.

\bibitem[FukSh]{FukSh}
M. Fukushima and T. Shima, 
On a spectral analysis for the Sierpinski gasket, \emph{Potential Analysis} \textbf{1} (1992), 1--35.

\bibitem[GarSt]{GarSt}
R. Garunktis and J. Steuding,
On the roots of the equation $\zeta(s)=\alpha$, e-print, arXiv:1011.5339 [mathNT], 2010.

\bibitem[Ger]{Ger}
J. Gerling,
\emph{Untersuchungen zur Theorie von Weyl-Berry-Lapidus,}
Graduate thesis (diplomarbeit), Dept. of Physics, Universit\"at Osnabr\"uck, Germany, May 1992.

\bibitem[GerScm1]{GerScm1}
J. Gerling and H.-J. Schmidt,
Self-similar drums and generalized Weierstrass functions, \emph{Physica A} \textbf{191} (1992), 536--539.

\bibitem[GerScm2]{GerScm2}
J. Gerling and H.-J. Schmidt,
Three-term asymptotics of the spectrum of self-similar fractal drums, \emph{J. Math. Sci. Univ. Tokyo} \textbf{6} (1999), 101--126.


\bibitem[Go]{Go}
J. A. Goldstein, 
\emph{Semigroups of Linear Operators and Applications},
Oxford Science Publications, Oxford Mathematical Monographs, 
Oxford Univ. Press, Oxford and New York, 1985. 

\bibitem[Ha]{Ha}
M. Haase,
\emph{The Functional Calculus for Sectorial Operators},
Operator Theory: Advances and Applications, vol. 169, Birkh\"auser Verlag, Berlin, 2006.

\bibitem[Ham1]{Ham1}
B. M. Hambly,
Brownian motion on a random recursive Sierpinski gasket,
\emph{Ann. Probab.} \textbf{25} (1997), 1059--1102.

\bibitem[Ham2]{Ham2}
B. M. Hambly,
On the asymptotics of the eigenvalue counting function for random recursive Sierpinski gaskets,
\emph{Probab. Theory Related Fields} \textbf{117} (2000), 221--247.

\bibitem[HamLa]{HamLa}
B. M. Hambly and M. L. Lapidus,
Random fractal strings: their zeta functions, complex dimensions and spectral asymptotics,
\emph{Trans. Amer. Math}. Soc. No. 1, \textbf{358} (2006), 285--314.

\bibitem[HeLa]{HeLa}
C. Q. He and M. L. Lapidus, Generalized Minkowski content, spectrum of fractal drums,
fractal strings and the Riemann zeta-function,
\emph{Memoirs Amer. Math. Soc}. No. 608,
\textbf{127} (1997), 1--97.

\bibitem[HerLa1]{HerLa1}
H. Herichi and M. L. Lapidus,
Fractal strings, the spectral operator and the Riemann hypothesis: Zeta values, Riemann zeroes, universality and phase transitions, memoir, preprint, 2012.

\bibitem[HerLa2]{HerLa2}
H. Herichi and M. L. Lapidus,
Convergence of the Euler product of the spectral operator in the critical strip,
in preparation, 2012.

\bibitem[HiPh]{HiPh}
E. Hille and R. S. Phillips,
\emph{Functional Analysis and Semi-groups}, 
Amer. Math. Soc. Colloq. Publ., vol. XXXI, rev. ed., Amer. Math. Soc., R.I., 1957.

\bibitem[Ing]{Ing}
A. E. Ingham,
\emph{The Distribution of Prime Numbers},
2nd ed. (reprinted from the 1932 ed.), Cambridge Univ. Press, Cambridge, 1992.

\bibitem[Ivi]{Ivi}
A. Ivic,
\emph{The Riemann Zeta}-\emph{Function: The theory of the Riemann zeta-function with applications}, John Wiley $\And$ Sons, New York, 1985.

\bibitem[JoLa]{JoLa}
G. W. Johnson and M. L. Lapidus,
\emph{The Feynman Integral and Feynman's Operational Calculus}, Oxford Science Publications, Oxford Mathematical Monographs, Oxford Univ. Press, Oxford and New York, 2000. (Paperback edition and corrected reprinting, 2002.)

\bibitem[KarVo]{KarVo}
A. A. Karatsuba and S. M. Voronin,
\emph{The Riemann  Zeta-Function}, 
Expositions in Mathematics, Walter de Gruyter, Berlin, 1992.

\bibitem[Kat]{Kat}
T. Kato,
\emph{Perturbation Theory for Linear Operators},
Springer Verlag, New York, 1995.

\bibitem[Ki]{Ki}
J. Kigami,
\emph{Analysis on Fractals}, Cambridge Univ. Press, Cambridge, 2001.

\bibitem[KiLa1]{KiLa1}
J. Kigami and M. L. Lapidus,
Weyl's problem for the spectral distribution of Laplacians on p.c.f self-similar fractals, \emph{Commun. Math. Phys}. \textbf{158} (1993), 93--125.

\bibitem[KiLa2]{KiLa2}
J. Kigami and M. L. Lapidus,
Self-similarity of volume measures for Laplacians on p.c.f self-similar fractals, \emph{Commun. Math. Phys}. \textbf{217} (2001), 165--180.

\bibitem[La1]{La1}
M. L. Lapidus, Fractal drum, inverse spectral problems for elliptic operators and a partial resolution of the Weyl--Berry conjecture, \emph{Trans. Amer. Math. Soc}.
\textbf{325} (1991), 465--529.

\bibitem[La2]{La2}
M. L. Lapidus,
Spectral and fractal geometry: From the Weyl--Berry conjecture for the vibrations of fractal drums to the Riemann zeta-function, in:
\emph{Differential Equations and Mathematical Physics}
(C. Bennewitz, ed.), Proc. Fourth UAB Internat. Conf.
(Birmingham, March 1990), Academic Press, New York, 1992, pp. 151--182.

\bibitem[La3]{La3}
M. L. Lapidus,Vibrations of fractal drums, the Riemann hypothesis,
waves in fractal media,\,and the Weyl--Berry conjecture, in:
\emph{Ordinary and Partial Differential Equations}
(B. D. Sleeman and R. J. Jarvis, eds.), vol. IV, Proc. Twelfth Internat. Conf. (Dundee, Scotland, UK, June 1992), Pitman Research Notes in Math. Series, vol. 289, Longman Scientific and Technical, London, 1993, pp. 126--209.

\bibitem[La4]{La4}
M. L. Lapidus, Fractals and vibrations: Can you hear the shape of a fractal drum?, \emph{Fractals},
No. 4, \textbf{3} (1995), 725--736.
(Special issue in honor of Benoit B. Mandelbrot's 70th birthday.) 

\bibitem[La5]{La5}
M. L. Lapidus,
\emph{In Search of the Riemann Zeros: Strings, fractal membranes and noncommutative spacetimes},
Amer. Math. Soc., Providence, R.I., 2008.


\bibitem[LaMa1]{LaMa1}
M. L. Lapidus and H. Maier,
Hypoth\`ese de Riemann, cordes fractales vibrantes et conjecture de Weyl--Berry modifi\'ee, \emph{C. R. Acad. Sci. Paris S\'er}. I \emph{Math}. \textbf{313} (1991), 19--24.

\bibitem[LaMa2]{LaMa2}
M. L. Lapidus and H. Maier, The Riemann hypothesis and inverse spectral problems for fractal strings, \emph{J. London Math. Soc. $($2$)$} \textbf{52} (1995), 15--34.

\bibitem[LaPe]{LaPe}
M. L. Lapidus and E. P. J Pearse,
Tube Formulas and complex dimensions of self-similar tilings, \emph{Acta Applicandae Mathematicae}
No. 1, \textbf{112} (2010), 91--137. (E-print: arXiv: math.DS/0605527v5, 2010; Springer Open Access: DOI 10.1007/S10440-010--9562-x.)

\bibitem[LaPeWi]{LaPeWi}
M. L. Lapidus, E. P. J Pearse and S. Winter,
Pointwise tube formulas for fractal sprays and self-similar tilings with arbitrary generators, \emph{Adv. Math.} \textbf{227} (2011), 1349--1398. (E-print: arXiv: 1006.3807v3 [math.MG], 2011.)


\bibitem[LaPo1]{LaPo1}
M. L. Lapidus and C. Pomerance,
Fonction z\^eta de Riemann et conjecture de Weyl--Berry pour les tambours fractals, \emph{C. R. Acad. Sci. Paris S\'er. I Math}. \textbf{310} (1990), 343--348.

\bibitem[LaPo2]{LaPo2}
M. L. Lapidus and C. Pomerance,
The Riemann zeta-function and the one-dimensional Weyl--Berry conjecture for fractal drums, \emph{Proc. London Math. Soc}. (3) \textbf{66} (1993), 41--69.

\bibitem[LaPo3]{LaPo3}
M. L. Lapidus and C. Pomerance,
Counterexamples to the modified Weyl--Berry conjecture on fractal drums, \emph{Math. Proc. Cambridge Philos. Soc}. \textbf{119} (1996), 167--178.

\bibitem[Lau]{Lau}
A. Laurincikas,
\emph{Limit Theorems for the Riemann Zeta-Function}, Kluwer Academic Publishers, Dordrecht, 1996.

\bibitem[La-vF1]{La-vF1}
M. L. Lapidus and M. van Frankenhuijsen,
Complex dimensions of fractal strings and oscillatory phenomena in fractal geometry and arithmetic,\,in:\,\emph{Spectral Problems in Geometry and Arithmetic} (T. Branson, ed.), Contemporary Mathematics, vol. 237, Amer. Math. Soc., Providence, R. I., 1999, pp. 87--105.

\bibitem[La-vF2]{La-vF2}
M. L. Lapidus and M. van Frankenhuijsen,
\emph{Fractal Geometry and Number Theory: Complex dimensions of fractal strings and zeros of zeta functions}, Birkh\"auser, Boston, 2000.

\bibitem[La-vF3]{La-vF3}
M. L. Lapidus and M. van Frankenhuijsen,
\emph{Fractal Geometry, Complex Dimensions and Zeta Functions: Geometry and spectra of fractal strings},
Springer Monographs in Mathematics, Springer, New York, 2006. (Second rev. and enl. ed., 2012, in press.)

\bibitem[La-vF4]{La-vF4}
M. L. Lapidus and M. van Frankenhuijsen (eds.), 
\emph{Fractal Geometry and Applications: A Jubilee of Benoit Mandelbrot}, Proc. Symp. Pure Math., vol. 72, Parts 1 \& 2, Amer. Math Soc., Providence, R.I., 2004.

\bibitem[Man]{Man}
B. B. Mandelbrot, \emph{The Fractal Geometry of Nature}, rev. and enl. ed (of the 1977 ed.), W. H. Freeman, New York, 1983.

\bibitem[Mat]{Mat}
P. Mattila,
\emph{Geometry of Sets and Measures in Euclidean Spaces: Fractals and rectifiability},
Cambridge Univ. Press, Cambridge, 1995.

\bibitem[Pat]{Pat}
S. J. Patterson, \emph{An Introduction to the Theory of the Riemann Zeta-Function}, Cambridge Univ. Press, Cambridge, 1988.

\bibitem[Paz]{Paz}
A. Pazy,
\emph{Semigroups of Linear Operators and Applications to Partial Differential Equations},
Springer-Verlag,
Berlin and New York, 1983.

\bibitem[Ram]{Ram}
R. Rammal,
Spectrum of harmonic excitations on fractals, \emph{J. de Physique} \textbf{45} (1984), 191--206.

\bibitem[RamTo]{RamTo}
R. Rammal and G. Toulouse, Random walks on fractal structures and percolation cluster, \emph{J. Physique Lettres} \textbf{44} (1983), L13--L22.

\bibitem[ReSi]{ReSi}
M. Reed and B. Simon, \emph{Methods of Modern  Mathematical Physics}, vol. I, \emph{Functional Analysis}, rev. and enl. ed. (of the 1975 ed.) and vol. II, \emph{Fourier Analysis, Self-Adjointness}, Academic Press, New York, 1980 and 1975.


\bibitem[Ru]{Ru}
W. Rudin, \emph{Functional Analysis}, 2nd ed. (of the 1973 ed.), McGraw-Hill, New York, 1991.

\bibitem[Sab1]{Sab1}
C. Sabot, Integrated density of states of self-similar Sturm--Liouville operators and holomorphic dynamics in higher dimension, \emph{Ann. Inst. H. Poincar\'e Probab. Statist.} \textbf{37} (2001), 275--311.

\bibitem[Sab2]{Sab2}
C. Sabot, Spectral analysis of a self-similar Sturm--Liouville operator, \emph{Indiana Univ. Math. J.} \textbf{54} (2005), 645--668.

\bibitem[Sab3]{Sab3}
C. Sabot, \emph{Spectral properties of self-similar lattices and iteration of rational maps}, M\'emoirs Soc. Math. France (New Series), No. 92, (2003), 1--104.

\bibitem[Sh]{Sh}
T. Shima, 
On eigenvalue problems for Laplacians on p.c.f. self-similar sets, \emph{Japan J. Indust. Appl. Math.} \textbf{13} (1996), 1--23.

\bibitem[Sc]{Sc}
M. Schechter,
\emph{Operator Methods in Quantum Mechanics}, Dover Publications, 2003.

\bibitem[Schw]{Schw}
L. Schwartz, \emph{Th\'eorie des Distributions}, rev. and enl. ed. (of the 1951 ed.), Hermann, Paris, 1996.

\bibitem[St]{St}
J. Steuding,
\emph{Value}-\emph{Distribution and $L$-Functions}, 
Lecture Notes in Mathematics, vol. 1877, 
Springer, Berlin, 2007.

\bibitem[Str]{Str}
R. S. Strichartz,
\emph{Differential Equations on Fractals: A Tutorial},
Princeton Univ. Press, 2006.

\bibitem[Tep1]{Tep1}
A. Teplyaev,
\emph{Spectral zeta function of symmetric fractals}, in:\,\emph{Fractal Geometry and Stochastics III}, Progress in Probability, vol. 57, Birkh\"auser-Verlag, Basel, 2004, pp. 245--262.

\bibitem[Tep2]{Tep2}
A. Teplyaev, Spectral zeta functions of fractals and the complex dynamics of polynomials, \emph{Trans. Amer. Math. Soc.} \textbf{359} (2007), 4339--4358.


\bibitem[Tit]{Tit}
E. C. Titchmarsh,
\emph{The Theory of the Riemann Zeta}-\emph{Function},
2nd ed. (revised by D. R. Heath-Brown), Oxford Science Publications, Oxford Mathematical Monographs, Oxford Univ. Press, Oxford, 1986. 


\bibitem[vB-Gi]{vB-Gi}
M. van den Berg and P. B. Gilkey,
A comparison estimate for the heat equation with an application to the heat content of the $s$-adic von Koch snowflake, \emph{Bull. London Math. Soc.} \textbf{30} (1998), 404--412. 
\end{thebibliography}
\end{document}